\documentclass[aps,prl,reprint,superscriptaddress,longbibliography,floatfix,maxcitenames=50]{revtex4-2}

\usepackage{amsmath,graphicx,epstopdf,latexsym,float}
\usepackage{xcolor}
\usepackage{amsfonts} 
\usepackage[colorlinks=false,bookmarks=false,citecolor=blue,linkcolor=blue,urlcolor=blue]{hyperref}
\usepackage[utf8]{inputenc}
\usepackage{amssymb}
\usepackage{tabularx}
\usepackage{braket}
\usepackage{nicefrac}
\usepackage{soul}

\begin{document}
\title{Single Atom Magnets on Thermally Stable Adsorption Sites: Dy on NaCl(100)}
\date{\today}
\author{M. Pivetta}
\email[Contact author: ]{marina.pivetta@epfl.ch}
\affiliation{Institute of Physics, Ecole Polytechnique F{\'e}d{\'e}rale de Lausanne (EPFL), CH-1015 Lausanne, Switzerland}
\author{M. Blanco-Rey}
\email[Contact author: ]{maria.blanco@ehu.es}
\affiliation{Departamento de Pol\'{\i}meros y Materiales Avanzados:  F\'{\i}sica, Qu\'{\i}mica y Tecnolog\'{\i}a, Facultad de Qu\'{\i}mica UPV/EHU, Apartado 1072, 20080 Donostia-San Sebasti\'an, Spain}
\affiliation{Donostia International Physics Center, Paseo Manuel de Lardiz\'abal 4, 20018 Donostia-San Sebasti\'an, Spain}
\affiliation{Centro de F\'{\i}sica de Materiales CFM/MPC (CSIC-UPV/EHU), Paseo Manuel de Lardiz\'abal 5, 20018 Donostia-San Sebasti\'an, Spain}
\author{S. Reynaud}
\author{R. Baltic}
\author{A. Rary-Zinque}
\author{S.~Toda~Cosi}
\author{F. Patthey}
\affiliation{Institute of Physics, Ecole Polytechnique F{\'e}d{\'e}rale de Lausanne (EPFL), CH-1015 Lausanne, Switzerland}
\author{B. V. Sorokin}
\affiliation{Institute of Physics, Ecole Polytechnique F{\'e}d{\'e}rale de Lausanne (EPFL), CH-1015 Lausanne, Switzerland}
\affiliation{Paul Scherrer Institut PSI, CH-5232 Villigen, Switzerland}
\author{A. Singha}
\affiliation{Institute of Physics, Ecole Polytechnique F{\'e}d{\'e}rale de Lausanne (EPFL), CH-1015 Lausanne, Switzerland}
\affiliation{Max Planck Institute for Solid State Research, Stuttgart
70569, Germany}
\author{F. Donati}
\affiliation{Institute of Physics, Ecole Polytechnique F{\'e}d{\'e}rale de Lausanne (EPFL), CH-1015 Lausanne, Switzerland}
\affiliation{Center for Quantum Nanoscience, Institute for Basic Science (IBS), Seoul 03760, Republic of Korea}
\affiliation{Department of Physics, Ewha Womans University, Seoul 03760, Republic of Korea}
\author{A. Barla}
\affiliation{Istituto di Struttura della Materia (ISM), Consiglio Nazionale delle Ricerche (CNR),
Trieste I-34149, Italy}
\author{L.~Persichetti}
\affiliation{Department of Materials, ETH Zurich, CH-8093 Zurich, Switzerland}
\affiliation{Dipartimento di Fisica, Universit\`a  di Roma “Tor Vergata”, I-00133 Roma, Italy}
\author{P. Gambardella}
\affiliation{Department of Materials, ETH Zurich, CH-8093 Zurich, Switzerland}
\author{A. Arnau}
\affiliation{Departamento de Pol\'{\i}meros y Materiales Avanzados:  F\'{\i}sica, Qu\'{\i}mica y Tecnolog\'{\i}a, Facultad de Qu\'{\i}mica UPV/EHU, Apartado 1072, 20080 Donostia-San Sebasti\'an, Spain}
\affiliation{Centro de F\'{\i}sica de Materiales CFM/MPC (CSIC-UPV/EHU), Paseo Manuel de Lardiz\'abal 5, 20018 Donostia-San Sebasti\'an, Spain}
\affiliation{Donostia International Physics Center, Paseo Manuel de Lardiz\'abal 4, 20018 Donostia-San Sebasti\'an, Spain}
\author{F. Delgado}
\affiliation{Departamento de F\'{\i}sica, Instituto Universitario de Estudios Avanzados en F\'{\i}sica At\'omica, Molecular y Fot\'onica (IUDEA), Universidad de La Laguna 38203, Tenerife, Spain}
\author{S. Rusponi}
\author{H. Brune}
\affiliation{Institute of Physics, Ecole Polytechnique F{\'e}d{\'e}rale de Lausanne (EPFL), CH-1015 Lausanne, Switzerland}

\begin{abstract}
We report magnetic bistability in single Dy atoms on NaCl(100) thin films. Individual Dy atoms substituting Na at the surface of the NaCl layer are thermally stable up to at least 300\,K, display $4f^{9}$ occupancy, out-of-plane easy magnetization axis, and long spin relaxation time $T_1$ of about 10\,s at 2.5\,K; thereby they are the first single atom magnet on a thermally stable adsorption site. Dy atoms adsorbed onto the Cl and bridge sites display $4f^{10}$ occupancy. Dy on top-Cl exhibit magnetic hysteresis and a $T_1$ of 550\,s at 0.3\,T and 2.5\,K. The observed slow magnetic relaxation of Dy on both adsorption sites introduces NaCl as an effective  platform for single atom magnets. 
\end{abstract}

\maketitle
\textit{Introduction.}\textemdash  Research on rare earth (RE) single-atom magnets (SAMs)~\cite{don16, bal16, nat17, mar17, all17, nat18, for19, don20, sin21, don21, bel22, sor23} has made progress in increasing the thermal stability of the magnetization and the associated height of the magnetic anisotropy barrier~\cite{nat18, sin21}. However, further progress is limited by the poor stability of the atoms on their adsorption sites with respect to the onset temperature of diffusion. Currently, the magnetically most stable SAMs are Ho atoms adsorbed on the oxygen site of MgO thin films, exhibiting magnetic relaxation times $T_1\,= 60$\,s at $T = 45$\,K~\cite{nat18}. However, their thermal diffusion from the top-O to the bridge site, where they lose their magnetic stability and become paramagnetic, sets in between 58 and 70\,K~\cite{fer17thesis}. For Dy atoms on the O sites of MgO, the magnetization is expected to be even more stable than that of Ho due to a 40\,\% higher anisotropy barrier~\cite{sin21,don21APL}, yet thermal diffusion to bridge sites may set in at even lower temperatures~\cite{fer17thesis, sin21}. Therefore, further improvement requires the creation of new systems in which the atom adsorption site is thermally stable up to higher temperatures. 

Embedding the atoms into the surface of the decoupling layer is expected to increase the diffusion barrier. It has been shown that some transition metal atoms replace Na or Cl at the surface of NaCl thin films~\cite{li14, li15, bib:li22}. For these atoms, however, one cannot expect very long $T_1$  because of the delocalized nature of the $d$ orbitals carrying the magnetic moment. Yet, NaCl is an efficient decoupling layer from a metallic support, allowing for example to control the charge of single atoms and molecules~\cite{rep04, ols07, leo11, li16}, or to access the intrinsic electronic  properties of molecules~\cite{rep05b, ros09, ima18, kai19, pat19}. Thus, NaCl thin films are promising candidates as SAM supports, although stabilization of the magnetization of single atoms imposes even stricter requirements.

Here we show that, by choosing the deposition temperature and the NaCl film thickness, we steer the adsorption of Dy onto top-Cl, bridge, or Na-substitutional sites. These different species allow us to compare the magnetic stability of highly-coordinated substitutional atoms with that of adatoms on the same decoupling layer. Using x-ray absorption spectroscopy (XAS) and x-ray magnetic circular dichroism (XMCD)~\cite{pia12}, we demonstrate that individual Dy atoms on NaCl thin films exhibit long $T_1\approx 10$\,s at 2.5\,K on Na-substitutional sites, where they are stable against diffusion up to room temperature (RT). Dy adatoms on top-Cl sites have an even longer spin relaxation time of $T_1 = 550$~s at 0.3\,T and 2.5\,K, although with limited structural stability. Note that the $T_1$ values measured with XMCD may underestimate the intrinsic relaxation times due to the influence of secondary electrons generated during x-ray absorption measurements~\cite{dre14, don16, don20}.  
\begin{figure*}[ht]
\centering
\includegraphics[width=1\textwidth]{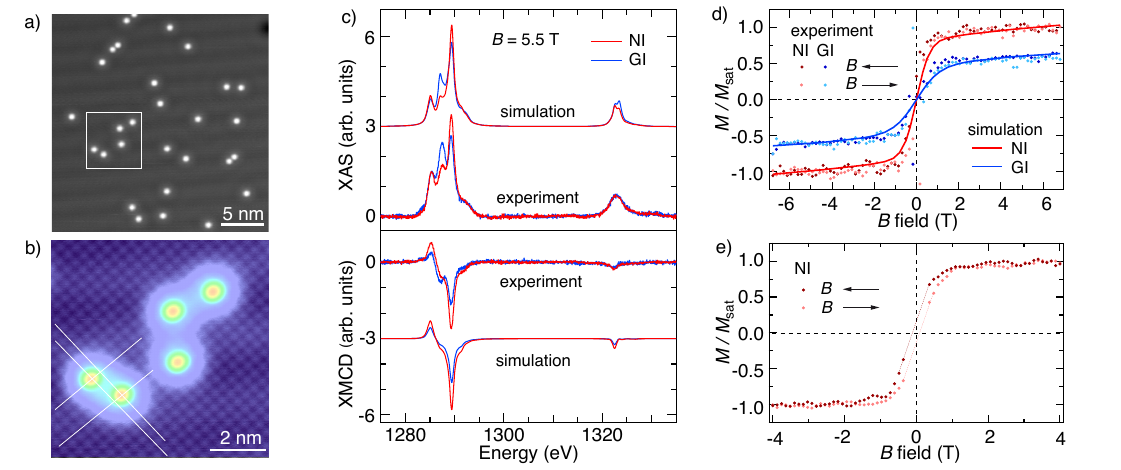}
\caption{a) STM images of Na-substitutional Dy atoms on 2\,ML NaCl/Ag(111) (Dy deposition temperature $T_{\rm dep}= 300$\,K,  Dy coverage $\theta_{\rm Dy}= 0.5$\,\%\,ML), the quasi-horizontal stripes are a moir\'e pattern of the NaCl ($V_{\rm t}=+300$\,mV, $I_{\rm t}=50$\,pA, $T_{\rm STM}= 6$\,K). b) Zoom into the region in the square; Cl atoms are imaged as protrusions, Dy atoms are in registry with Na sites ($V_{\rm t}=-50$\,mV, $I_{\rm t}=50$\,pA). 
 c) Experimental and simulated XAS and XMCD spectra for Na-substitutional Dy atoms on 3.5~ML NaCl/Ag(111) ($T_{\rm dep}= 300$\,K, $\theta_{\rm Dy}=1.5$\,\%\,ML), at the Dy $M_{4,5}$ edges ($B = 5.5$\,T, $T_{\rm XAS} = 2.5$\,K), vertically shifted for clarity. d,e) Magnetization curves acquired by recording the XMCD signal at 1289.3\,eV with field sweep rate $|\dot B | = 2$\,T\,/min: d) normal (NI) and grazing (GI) incidence curves, compared with simulated ones, at flux $\phi = 1.1 \cdot 10^{-2}$\,ph\,nm$^{-2}$\,s$^{-1}$ (the amplitude of the simulated curves at large fields is normalized to the amplitude of the experimental ones) and e) NI curve showing opening at a lower photon flux of $\phi = 7.5 \cdot 10^{-3}$\,ph\,nm$^{-2}$\,s$^{-1}$.  Measured and simulated magnetization curves are at $T= 2.5$\,K.
\label{subst}
}
\end{figure*}
These findings demonstrate that NaCl is an effective supporting layer for SAMs, as it protects the atomic spin from electron and phonon scattering and provides adequate crystal fields (CF) promoting slow spin relaxation.

\textit{Substitutional Dy atoms.}\textemdash  
Upon RT deposition, Dy atoms substitute Na atoms at the surface of the NaCl layer, as verified by STM for $2-3$\,ML NaCl films (see Supplemental Material~\cite{SM} for more details). The STM image shown in Fig.~\ref{subst}(a), acquired on 2\,ML NaCl grown on Ag(111), exhibits isolated protrusions with an apparent height of $(100 \pm 5)$\,pm, which is unusually shallow for a rare-earth (RE) adatom~\cite{bal16, fer17, fer17thesis, sin21}. The uniform apparent height indicates a unique substitutional site, as reported for Co on NaCl/Au~\cite{li15,bib:li22}, while coexistence of several sites was found for Cr~\cite{li14}. The atomic resolution image in Fig.~\ref{subst}(b) shows Cl atoms as protrusions~\cite{heb99, rep01, ols05, li14, li15b} and thereby reveals that the Dy atoms are at Na sites. Our density functional theory (DFT) calculations show that, at the surface of 2\,ML thick NaCl films on Ag, Na-vacancy formation followed by replenishment with a RE atom is an exothermic process ~\cite{SM}. Combining the STM and DFT results, we conclude that the Dy atoms substitute Na atoms in the topmost layer of NaCl. Their thermal stability is inferred from the observation of isolated atoms, which excludes clustering, as well as from STM and DFT results indicating absence of diffusion at RT~\cite{SM}.

\begin{figure*}[ht!]
\centering
\includegraphics[width=1\textwidth]{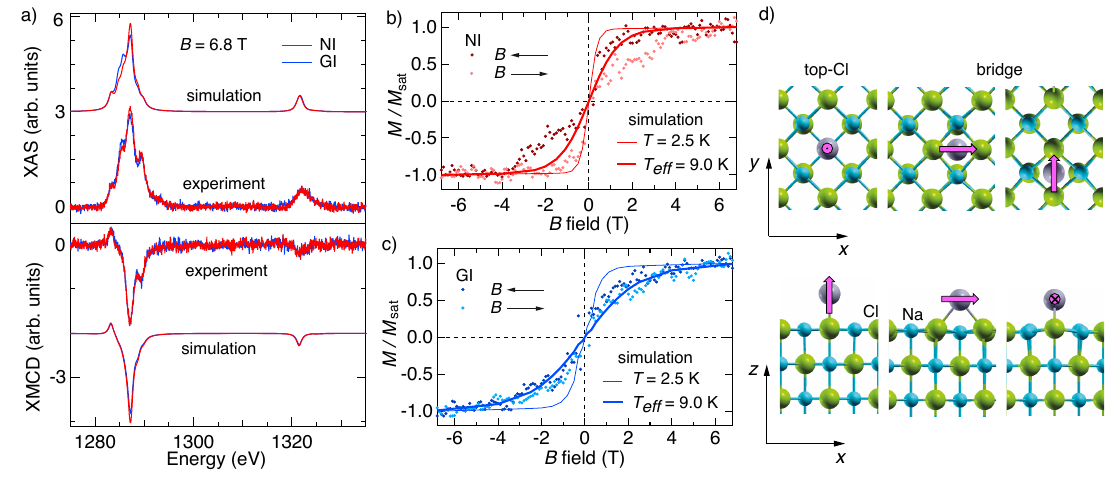}
\caption{a) Experimental and simulated XAS and XMCD spectra at the Dy $M_{4,5}$ edges of Dy adatoms on 8\,ML NaCl/Cu(111) ($T_{\rm dep}= 4$\,K,  $\theta_{\rm {Dy}}=1.9$\,\%\,ML, $B=6.8$\,T, $T_{\rm XAS}=2.5$\,K).
b) NI and c) GI magnetization curves acquired by recording the XMCD signal at 1287.2\,eV ($|\dot B | = 2$\,T\,/min, $\phi =1.5\cdot10^{-2}$\,ph\,nm$^{-2}$\,s$^{-1}$), and simulated equilibrium curves at $T=2.5$\,K and $T_{\rm eff}=9.0$\,K. d) Top-view and side-view of Dy adatoms at top-Cl and two symmetry-equivalent bridge adsorption sites; pink arrows indicate the easy magnetization directions for the different species.
 }
\label{adat}
\end{figure*}

To investigate the magnetic properties of the substitutional Dy atoms, we have carried out x-ray absorption measurements on a sample prepared by RT Dy deposition on a NaCl film of 3.5\,ML average thickness grown on Ag(111). This NaCl coverage ensures that there is no bare Ag exposed and a layer thicknesses that is still close to the one investigated by STM. The XAS spectra shown in Fig.~\ref{subst}(c) reveal that Dy in Na-substitutional sites is in the $4f^{9}$ electronic configuration, compared to $4f^{10}$ in the gas phase~\cite{sin17, don21}. This behavior is expected in light of the high coordination of the Dy atoms in this geometry~\cite{bal18, joh79, bre93, pet15}. The higher XMCD intensity at normal incidence (NI, $\theta=0^{\circ}$) compared to grazing (GI, $\theta=60^{\circ}$) indicates an out-of-plane easy magnetization axis. The magnetization curves shown in Fig.~\ref{subst}(d) confirm this anisotropy. Reducing the photon flux ($\phi$), see Fig.~\ref{subst}(e), we observe a narrow opening, extending up to about 1\,T. Given the field sweep rate of 2\,T/min and the maximum hysteresis opening of $\approx 0.25$\,T observed at $\approx 0.3$\,T, we estimate $T_1 \approx 10$\,s, that can be potentially longer at even lower $\phi$. 
Since during fast field sweeps the XMCD signal close to zero field is strongly affected by noise, a firm conclusion about magnetic remanence cannot be drawn. However, resonant quantum tunneling of the magnetization (QTM) at zero field is ruled out for the Dy $4f^9$ Kramers spin. In the simplest scenario, supported by the constant low-field  magnetic susceptibility observed in Fig.~\ref{subst}(d), a linear dependence of the magnetization in the low-field range is expected, as indicated by the dotted lines in Fig.~\ref{subst}(e). Considering such behavior in the range $|B| < 0.4$\,T, we deduce a remanence of about 25\,\% of the saturation magnetization, $M_{\rm sat}$, and a coercive field of about $\pm0.1$\,T. Thus, Na-substitutional Dy atoms on NaCl are the first SAMs to have a thermally stable adsorption site as well as lateral coordination.

XAS and XMCD spectra at NI and GI, x-ray linear dichroism (XLD) spectra \cite{SM}, as well as equilibrium magnetization curves are simulated using electronic multiplet calculations with the CF modeled by point-charges. Position and value of these charges are derived from our DFT calculations (see End Matter and \cite{SM}). Our  calculations reproduce all the spectral features well and provide the magnetic level scheme for the Dy atoms. 
The resulting magnetization reversal paths will be discussed below in the section \textit{Magnetic relaxation}.

\textit{Dy adatoms.}\textemdash 
Our STM investigations show that Dy deposition at low temperature ($T_{\rm dep}=10$\,K) on $2 - 3$\,ML NaCl results in the coexistence of two species of Dy atoms: adatoms on bridge sites and substitutional atoms (LT-substitutional)~\cite{SM}.
The proportion of LT-substitutional atoms \textit{vs.} adatoms decreases for increasing NaCl thickness. Therefore, in order to maximize the adatom population and be able to investigate their magnetic properties in x-ray absorption experiments, we used thicker NaCl films ($\approx 8$\,ML). The XAS spectra in Fig.~\ref{adat}(a) reveal a $4f^{10}$ occupancy~\cite{bal16, sin17, don21}; the shoulder at 1289.3\,eV, indicating some $4f^{9}$ contribution, is attributed mainly to contaminated Dy since its intensity increases with time~\cite{bal16, bal18}, although a small contribution of LT-substitutional atoms cannot be excluded. The XMCD spectra do not show significant angular dependence. 

\begin{figure*}[ht!]
\centering
\includegraphics[width=1\textwidth]{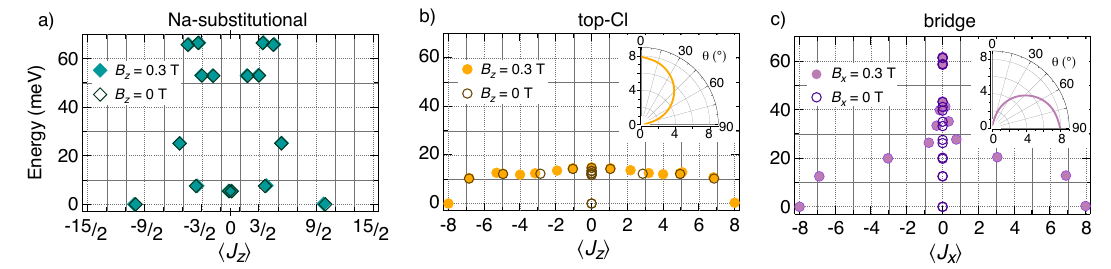}
\caption{Magnetic level schemes at $B=0.3$\,T and $B=0$\,T provided by the multiplet calculations for a) Na-substitutional Dy atoms, b) top-Cl and c) bridge Dy adatoms. Insets in b) and c) show the polar dependence of the total angular momentum component in a magnetic field of 6.8\,T  along the respective easy axes.
} 
\label{energy}
\end{figure*}

The similar intensity of the XMCD spectra acquired in NI and GI is rationalized by the different easy magnetization axis of adatoms at different adsorption sites, namely, top-Cl and bridge species, as sketched in Fig.~\ref{adat}(d). Our STM characterization, performed  on $2 – 3$\,ML thick NaCl islands, reveals adatoms on bridge adsorption sites only~\cite{SM}. However, experiments on Ho and Dy on various oxides show increasing adsorption on top-O sites with increasing film thickness~\cite{don21, bel22, sor23}. Similar to this behavior on oxides, our DFT results show that the top-Cl is the preferred site for 6\,ML NaCl~\cite{SM}. Therefore, on a NaCl film with an average thickness of 8\,ML, we expect the top-Cl species to coexist with the bridge species observed for lower NaCl thickness. 
NaCl grows on (111) surfaces with many different azimuthal orientations~\cite{ben99}, each creating two bridge sites rotated by 90$^\circ$. This leads to a multiplicity of in-plane bridge site orientations that we simulate for the GI geometry by an average bridge site oriented at 45$^\circ$ with respect to the field and beam direction. In this way, we reproduce the XAS, XMCD, and XLD~\cite{SM} spectra, as well as the isotropic behavior of the ensemble with an abundance of 60\,\% top-Cl and 40\,\% bridge adatoms. 
Finally, the magnetization curves in Fig.~\ref{adat}(b,c) show hysteretic behavior, with opening visible up to $\approx 4$\,T in NI, reflecting long spin lifetimes but presumably without magnetic remanence.  

\textit{Magnetic relaxation.}\textemdash  
For perfectly axial systems (C$_{\infty v}$ symmetry), such as free-standing diatomic lanthanide compounds~\cite{ung11, zha20}, the  eigenstates are given by the $(2J_{4f}+1)$ values of $J_z$, and therefore one needs to overcome the full anisotropy barrier to reverse the magnetic moment. More realistic systems include both axial and lateral (transverse) ligands, corresponding to a CF of reduced symmetry. For example, for Dy on the top-Cl site, because of the four  next-nearest neighbor Na ions in the surface plane, the CF has a fourfold symmetry (C$_{4v}$), while for Dy on the bridge site the CF symmetry reduces to C$_{2v}$. In a C$_{nv}$ CF, each eigenstate is a linear combination of pure $J_z$ states differing by $n$ (mixed state)~\cite{hub14}. Depending on the CF symmetry and $J_{4f}$, this mixing can lead to splitting of the otherwise degenerate states of the ground doublet, resulting in QTM, and/or of the excited doublets, resulting in thermally activated (TA) QTM ~\cite{fri96, tho96, gar97}. In TA-QTM, the energy scale plays a crucial role, since only populated doublets can participate in the reversal process. Usually, excited doublets are thermally populated by spin-phonon scattering, leading to thermalization of the isolated spin to the atom support (bath) temperature. In XAS experiments, excited states can also be occupied by scattering between the spin and the secondary electrons generated during x-ray absorption by the support. The energy distribution of these electrons differs from the equilibrium distrubution of the conduction electrons. Depending on the relative efficiency of the spin-phonon and spin-electron scattering, the isolated spin is at thermal equilibrium with the bath (spin-phonon dominates), or it is out of equilibrium, with a non-Boltzmann population of the states (spin-secondary electron dominates), since spin-phonon scattering is  not efficient enough to allow thermalization to the bath  (phonon bottleneck)~\cite{vanv41, sto65, ste67, gil75, chi00, bil21thesis}.  This non-thermal state occupation has two main consequences: a) reversal paths via excited doublets are active even at temperatures at which these states would not be populated, thus resulting in measured $T_1$ shorter than the intrinsic value \cite{don20}; b) the magnetic susceptibility is reduced with respect to the one expected at the measurement temperature.

For the Na-substitutional Dy atoms presented in Fig.~\ref{subst},  the multiplet simulations find $S_{4f} = \nicefrac{5}{2}$ and $L_{4f} = 5$, corresponding to a $J_{4f} = \nicefrac{15}{2}$ ground multiplet. The corresponding quantum level scheme shown in Fig.~\ref{energy}(a) helps us rationalize the origin of the observed, although narrow, magnetization curve hysteresis.
For this Kramers spin the levels of the ground-state doublet are protected against QTM at zero field, enabling magnetic remanence. The energy scheme and the wavefunction composition of the ground doublet support slow spin relaxation also at moderate field. 
For example at 0.3\,T, corresponding to the maximum hysteresis opening, the ground doublet wavefunctions have, with $|\pm \nicefrac{15}{2} \rangle$ (64.5\,\%) and $|\pm \nicefrac{7}{2} \rangle$ (12.5\,\%), most of the weight on one side of the barrier, and the remaining weight, with  $|\mp \nicefrac{1}{2} \rangle$ (20.5\,\%) and $|\mp \nicefrac{9}{2} \rangle$ (2.5\,\%), on the other side. In addition, the first excited doublet, with very low $\langle J_z \rangle$, is found at 5.4\,meV and is negligibly populated at $T=2.5$\,K. Both features imply that spin reversal requires scattering with secondary electrons generated by the x-ray absorption and/or phonons.

The magnetization curves simulated at the experimental temperature $T=2.5$\,K reproduce well the measured magnetic susceptibility, see Fig.~\ref{subst}(d), suggesting that spin-phonon scattering largely dominates. Consequently,  $T_1 \approx 10$\,s  estimated at the lowest flux is likely representative of the intrinsic spin relaxation time. 

For top-Cl Dy adatoms, the magnetic level scheme of the non-Kramers $J_{4f} = 8$  ($S_{4f} = 2$, $L_{4f} = 6$) ground multiplet is reported in Fig.~\ref{energy}(b). The atoms have strong out-of-plane easy magnetization axis, as seen in the polar plot  in the inset, with the ground state mainly given by the superposition of the $J_z  = \pm 8$ states induced by the C$_{4v}$ CF, with $J_z  = \pm 4, 0$ contributing less than 0.1\%.  This composition of the ground doublet leads to QTM in zero field, with an energy splitting $\Delta E = 0.65$\,neV,  explaining the absence of remanence. A similar ground doublet has been found for $4f^{10}$ top-O Dy on thick MgO~\cite{don21}, while in the case of Ho on MgO~\cite{sin21b}, showing magnetic remanence, the ground doublet splitting is negligible.

Bridge adatoms also have $J_{4f} = 8$, but the easy magnetization  axis is along the Cl-Cl direction ($x$); the C$_{2v}$ CF leads to states that are strongly mixed, as clearly displayed in Fig.~\ref{energy}(c). The ground doublet energy spitting in zero field is $\Delta E = 1.6$\,$\mu$eV, implying that at small fields $B_x \gtrsim 10$\,mT, the states of the ground doublet are the pure $J_z= 8$ and $J_z=-8$ states. In such fields and at the used experimental temperature, we expect long spin lifetime also for Dy adatoms at bridge sites, thanks to the large energy separation between ground and first excited doublet (12.5\,meV).

We use the level schemes of the adatoms to simulate the equilibrium magnetization curves, which lie within the upward and downward branches in presence of hysteresis. Assuming the experimental  temperature $T=2.5$\,K, the simulated NI and GI curves show a too large magnetic susceptibility for $|B| < 1$\,T, see Fig.~\ref{adat}(b,c), implying that the measurements are carried out in a regime where scattering between spin and secondary electrons is substantial. In order to account for this, we introduce an adatom effective temperature $T_{\rm eff}$ used as fit parameter. We reproduce the GI magnetization curve, which is close to equilibrium, with $T_{\rm eff}\approx 9$\,K. The same temperature describes well the equilibrium curve expected at NI, too. 
The fact that we need an effective temperature above the substrate temperature for the adatoms on thick NaCl films and not for the substitutional ones on thinner films can be rationalized by two qualitative arguments. First, we expect spin-phonon coupling to be less effective for a physisorbed top-Cl Dy atom sticking-out from the surface, than for a Na-substitutional atom embedded in the NaCl film. Second, scattering with the conduction electrons of the metal substrate as source for spin thermalization is suppressed for 8\,ML NaCl, while it plays a role for thinner NaCl films, similarly to the trend observed for Fe adatoms on MgO~\cite{pau17}.

The  measured spin relaxation time at NI, where only the top-Cl sites contribute to the signal as seen in the polar plots in Fig.~\ref{energy}, is  $T_1 = (550\ \pm\ 100)$\,s in the 0.1\,T to 0.5\,T range~\cite{SM}. A similar field-independent behavior has been reported for Ho top-O atoms on MgO~\cite{don20}, and is a signature of $T_1$ dominated by scattering with secondary electrons. Because these electrons have a very broad energy spectrum, all the spin states are populated and involved in the relaxation process, which therefore does not proceed via phonon-induced TA-QTM. Reducing the x-ray photon flux produces an increase in the observed $T_1$~\cite{don16}, implying that the measured value is a lower limit for the intrinsic lifetime. 

Note that the previous analysis is based on $J_{4f}$ only. However, magnetism in RE can also arise from the spin polarization of the external $5d6s$ shells. Their contribution to the total magnetic moment is small, but they can have a strong effect on the relaxation paths since the CF will mix states of the total angular momentum combining both $\boldsymbol{J}_{4f}$ and $\boldsymbol{J}_{5d6s}$, thus leading to different split doublets~\cite{cur23}.

Inelastic electron tunneling spectroscopy (IETS) allows to  reveal the $5d6s$ shells spin polarization: if present, inelastic excitations are observed at the energy corresponding to the reversal of $\boldsymbol{S}_{5d6s}$ coupled by intra-atomic exchange to $\boldsymbol{S}_{4f}$~\cite{piv20}. Since we do not observe any IETS features for the $4f^9$ substitutional Dy atoms, we deduce that their $5d6s$ shells are not spin-polarized. 
 
Concerning the $4f^{10}$ top-Cl site on thick NaCl films, DFT finds physisorption with a negligible spin polarization of the $5d6s$ shells as indicated by the small net Bader charge values~\cite{SM}. This prediction is in agreement with the shape of the magnetization curve shown in Fig.~\ref{adat}(b), demonstrating QTM via the ground doublet at zero field. 
Note that a very similar magnetization curve has been measured for $4f^{10}$ Dy at top-O sites on thick MgO films on Ag(100)~\cite{don21}, where DFT predicts an occupation of the $5d6s$ shells similar to the one we find for the thick NaCl case. 

\textit{Conclusion}\textemdash 
We have shown that both Na-substitutional and top-Cl Dy atoms on NaCl thin films display long spin lifetime. In particular, substitutional Dy represents the first system where single atoms are thermally stable up to 300\,K and display magnetic remanence at least at low temperature. We are able to explain these findings with the help of DFT calculations and multiplet simulations that reproduce the experimental data. The long $T_1$ observed for Dy on both sites demonstrates that NaCl thin films are suitable supporting layers for SAMs. This implies an efficient decoupling from the electrons of the metal substrate and a low local phonon density of states at the Dy sites at the relevant energies for magnetic relaxation~\cite{don20,bib:garai23,kra23}.

\begin{acknowledgments}
We acknowledge funding from the Swiss National Science Foundation 
(200021$\_$175941, 200021$\_$146715, 200020$\_$157081, 200020$\_$176932, and 200020$\_$204426). 
We acknowledge funding of project PRIN2022 "MAGNETISE" 2022KXN79M (Funded by the European Union - Next Generation EU, Mission 4 Component 1 CUP B53D23004280006).
We acknowledge funding of projects PID2022-137685NB-I00 and PID2022-138269NB-I00, funded by MCIN/AEI/10.13039/501100011033 and by ``ERDF A way of making Europe''; IT-1527-22, funded by the Department of Education, Universities and Research of the Basque Government. Computational resources were provided in part by the DIPC computing center.
\end{acknowledgments}


\appendix

\section{End Matter}

\textit{Charge density distribution, crystal field.}\textemdash  
We combine DFT and multiplet simulations to rationalize the magnetic properties observed for Dy at the different sites. The adsorption behavior of Dy species with $4f^9$ ($4f^{10}$) occupation is simulated by choosing the half-filled $f$-shell atom Gd (Eu). In this way, splitting of $f$ levels is neglected and the $f$ electrons are treated as core electrons. 
This is a reasonable approximation, as localized $f$ electrons do not take an active role in the binding to the substrate. The same rationale lies behind the so-called Yttrium-analogue approximation, successfully applied to CFs of RECo$_5$ intermetallics~\cite{bib:patrick19}, or using Gd instead of Y and enforcing Hund’s rule \cite{bib:lee25} The DFT calculations show that energetically preferred sites for both Eu and Gd are the Na-substitutional for the 2\,ML film, the bridge for 3\,ML and the top-Cl for 6\,ML. A projected density of states (PDOS) analysis of the Gd case allows us to identify a mechanism for transfer of the $5d$  electron to the lattice. As shown in the Supplemental Material Figs.~S9 and S10~\cite{SM}, the efficiency of this mechanism varies with the Gd adsorption site type and film thickness. Although our calculations cannot predict the $f^9$ or $f^{10}$ filling in Dy, this finding for Gd is compatible with the experimental observation of a site-dependent Dy $4f$ occupancy.

\begin{figure}[]
\centering
\includegraphics[width=1\columnwidth]{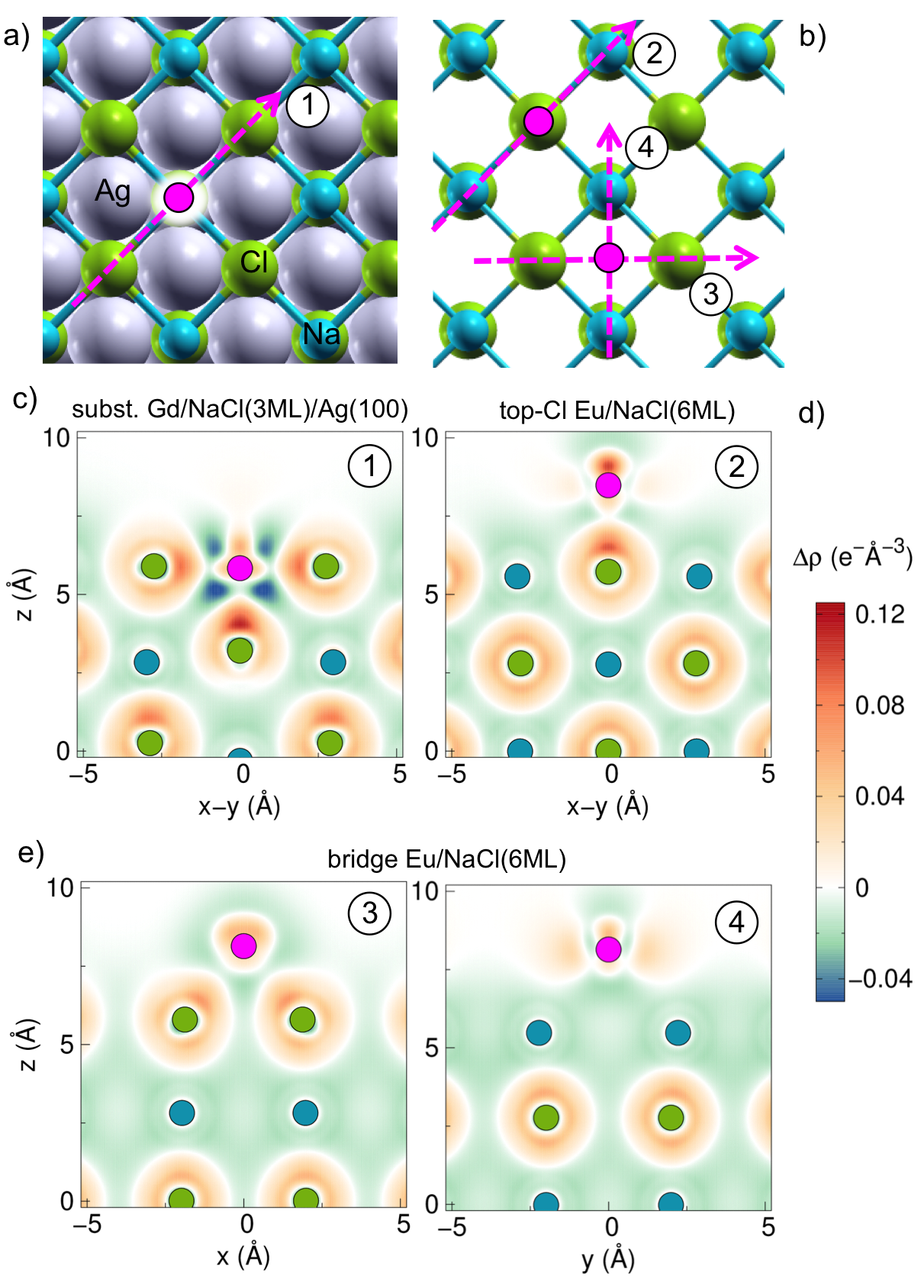}
\caption{Difference between the self-consistent DFT charge density and the superposition of individual atomic pseudo-charges~\cite{bib:kresse99} placed at the equilibrium atomic coordinates. a) Top view of the NaCl(3ML)/Ag(100) substrate, Gd  (magenta) occupies a Na-substitutional site. b) Top view of the NaCl(6ML) substrate, Eu (magenta) occupies a top-Cl or a bridge site. c-e) Transversal cuts of the charge density difference $\Delta \rho$ with respect to the superposition of atomic densities. The $z=0$ origin corresponds to the average height of the third NaCl layer. The horizontal distances are taken along the directions indicated by the dashed magenta arrows, labeled 1-4 in panels a) and b).} 
\label{dft}
\end{figure}

From DFT we also obtain the information to construct the point charge CF model used in the multiplet simulations.  Figure~\ref{dft} represents the interstitial electron density redistribution upon bond formation in the crystal. In all cases, the negative charge accumulates between the RE and the Cl atoms (red-orange), while the Na atoms remain mostly unperturbed. 

Inspection of Fig.~\ref{dft}(c) shows that for substitutional Gd atoms the CF consists of a dominant axial term, due to the excess of negative charge accumulated between the Gd and the Cl atom just beneath, combined with an intense transverse fourfold term resulting from the interaction with the four lateral nearest Cl atoms. 
Figure~\ref{dft}(d) shows that the  CF at the top-Cl site has mainly axial character, due to the Cl atom just beneath the Eu, with a small fourfold perturbation arising from the nearest Na atoms. Differently, at the bridge site shown in Fig.\ref{dft}(e), the stronger interaction is in the plane containing the two nearest Cl atoms, resulting in a twofold CF.

Values and positions of point negative (positive) charges are inferred based on the accumulation (depletion) of electron charge in the region surrounding the RE, using the approach applied to Dy and Ho adatoms on BaO thin films~\cite{sor23}. Charge values and positions are provided in the SM~\cite{SM}. Finally, multiplet simulations are carried out with the multiX software~\cite{uld12}. The CF built following this procedure not only explains the magnetism of Dy on NaCl, but also that of Ho~\cite{SM}, further supporting the use of DFT calculations with Gd and Eu as models. Indeed, we expect this result to extend to RE atoms with similar $f-d$ atomic correction energies (Nd, Dy, Ho and Er atoms have values in the 1.16-1.45\,eV range)~\cite{bib:delin97,sin17}. 

Note also that in our approach the results of the DFT calculations (charge values and spatial distributions) are used to determine the CF, with a minimum number of free parameters, namely, the positive charge of the Na ions and the charge of the axial Cl ion for substitutional Dy, and only the positive charge of the Na ions for Dy on top-Cl and bridge sites. No further adjustment is applied for a set of simulations which includes the XAS, XMCD, XLD spectra and the hysteresis curves, for the two orientation of the external field. This is a stringent approach compared to standard ones using a CF description via Stevens or Wybourne operators where at least five independent parameters are required. These are often adjusted to experimental data, as it is not trivial to relate them to the electronic structure details provided by DFT. In the light of these arguments, we consider the agreement between experiment and simulation remarkably good, with the discrepancy between experimental and simulated curves being small.

\makeatletter\@input{aux4_ms_16_SI.tex}\makeatother
 
\end{document}


\title{Supplemental Material \\Single Atom Magnets on Thermally Stable Adsorption Sites: Dy on NaCl(100)}
\date{\today}
\author{M. Pivetta}
\affiliation{Institute of Physics, Ecole Polytechnique F{\'e}d{\'e}rale de Lausanne (EPFL), CH-1015 Lausanne, Switzerland}
\author{M. Blanco-Rey}
\affiliation{Departamento de Pol\'{\i}meros y Materiales Avanzados:  F\'{\i}sica, Qu\'{\i}mica y Tecnolog\'{\i}a, Facultad de Qu\'{\i}mica UPV/EHU, Apartado 1072, 20080 Donostia-San Sebasti\'an, Spain}
\affiliation{Donostia International Physics Center, Paseo Manuel de Lardiz\'abal 4, 20018 Donostia-San Sebasti\'an, Spain}
\author{S. Reynaud}
\author{R. Baltic}
\author{A. Rary-Zinque}
\author{S.~Toda~Cosi}
\author{F. Patthey}
\affiliation{Institute of Physics, Ecole Polytechnique F{\'e}d{\'e}rale de Lausanne (EPFL), CH-1015 Lausanne, Switzerland}
\author{B. V. Sorokin}
\affiliation{Institute of Physics, Ecole Polytechnique F{\'e}d{\'e}rale de Lausanne (EPFL), CH-1015 Lausanne, Switzerland}
\affiliation{Paul Scherrer Institut PSI, CH-5232 Villigen, Switzerland}
\author{A. Singha}
\affiliation{Institute of Physics, Ecole Polytechnique F{\'e}d{\'e}rale de Lausanne (EPFL), CH-1015 Lausanne, Switzerland}
\affiliation{Max Planck Institute for Solid State Research, Stuttgart
70569, Germany}
\author{F. Donati}
\affiliation{Institute of Physics, Ecole Polytechnique F{\'e}d{\'e}rale de Lausanne (EPFL), CH-1015 Lausanne, Switzerland}
\affiliation{Center for Quantum Nanoscience, Institute for Basic Science (IBS), Seoul 03760, Republic of Korea}
\affiliation{Department of Physics, Ewha Womans University, Seoul 03760, Republic of Korea}
\author{A. Barla}
\affiliation{Istituto di Struttura della Materia (ISM), Consiglio Nazionale delle Ricerche (CNR),
Trieste I-34149, Italy}
\author{L.~Persichetti}
\affiliation{Department of Materials, ETH Zurich, CH-8093 Zurich, Switzerland}
\affiliation{Dipartimento di Fisica, Universit\`a  di Roma “Tor Vergata”, I-00133 Roma, Italy}
\author{P. Gambardella}
\affiliation{Department of Materials, ETH Zurich, CH-8093 Zurich, Switzerland}
\author{A. Arnau}
\affiliation{Centro de F\'{\i}sica de Materiales CFM/MPC (CSIC-UPV/EHU), Paseo Manuel de Lardiz\'abal 5, 20018 Donostia-San Sebasti\'an, Spain}
\affiliation{Departamento de Pol\'{\i}meros y Materiales Avanzados:  F\'{\i}sica, Qu\'{\i}mica y Tecnolog\'{\i}a, Facultad de Qu\'{\i}mica UPV/EHU, Apartado 1072, 20080 Donostia-San Sebasti\'an, Spain}
\affiliation{Donostia International Physics Center, Paseo Manuel de Lardiz\'abal 4, 20018 Donostia-San Sebasti\'an, Spain}
\author{F. Delgado}
\affiliation{Departamento de F\'{\i}sica, Instituto Universitario de Estudios Avanzados en F\'{\i}sica At\'omica, Molecular y Fot\'onica (IUDEA), Universidad de La Laguna 38203, Tenerife, Spain}
\author{S. Rusponi}
\author{H. Brune}
\affiliation{Institute of Physics, Ecole Polytechnique F{\'e}d{\'e}rale de Lausanne (EPFL), CH-1015 Lausanne, Switzerland}

\begin{abstract}

\end{abstract}

\maketitle

\setlength{\parskip}{3pt}
\renewcommand{\thefigure}{S\arabic{figure}}
\renewcommand{\thetable}{S\arabic{table}}

\section{Experimental}

\subsection{Sample preparation}

The Cu(111) and Ag(111) single crystals were prepared \textit{in situ} by repeated sputtering and   annealing cycles. NaCl thin films were grown from NaCl powder (99.5\,\% purity) using an effusion cell heated up to 770\,K, on the substrate kept at room temperature (RT). For the scanning tunneling microscopy (STM) experiments, we used a nominal coverage of the order of 0.5\,ML, resulting in regions of bare substrate coexisting with extended NaCl islands 2 or 3\,ML thick. For the synchrotron experiments, $2-10$\,ML of NaCl were grown; note that a minimum nominal coverage of $ 3-4$\,ML ensures the absence of exposed bare metal substrate. 

Rare earth (RE) atoms (Dy and Ho) were deposited from high purity (99.9\,\%) thoroughly degassed rods using \textit{e}-beam evaporators, on the substrate kept either at RT, or at approximately 10\,K (STM) and 4\,K (synchrotron experiments) for low temperature (LT) deposition. The RE coverage $\theta_{\mathrm RE}$ is given in monolayers (ML), with one ML being one RE atom per NaCl surface unit cell ($\approx 0.16$\,nm$^2$). For the STM experiments we used coverages between 0.5 and 1\,\%\,ML. For synchrotron experiments, RE coverage was below $\approx 2.5$\,\%\,ML in order to maximize the XAS signal while keeping the mean interatomic distance large enough to avoid magnetic interactions (see also Sec.~\ref{coupling}) and cluster formation.

\subsection{STM measurements}

Scanning tunneling microscopy measurements were performed with a home-built STM~\cite{gai92}. The images were recorded in constant current mode using electrochemically etched W or MnNi tips. The bias voltage $V_{\mathrm t}$ refers to the sample. All images were acquired at low temperature, $T_{\rm STM} = 6$\,K, except for the room temperature measurements described in Sec.\ref{Dysubstit}.

\subsection{X-ray absorption and dichroism measurements}

The x-ray absorption measurements were performed at the EPFL/PSI X-Treme beamline of the Swiss Light Source~\cite{pia12}. The experiments were carried out using both circularly ($\sigma^+$, $\sigma^-$) and linearly ($\sigma^v$, $\sigma^h$) polarized x-rays in the total electron yield (TEY) mode, at a sample temperature of 2.5\,K. The XAS corresponds to $(\sigma^+ + \sigma^-)$, while XMCD and XLD are defined as $(\sigma^+ - \sigma^-)$ and $(\sigma^h - \sigma^v)$, respectively. The measurements were carried out at the $M_{4,5}$ absorption edges of Dy and Ho. Mainly two incidence angles with respect to the surface normal have been used, namely normal incidence (NI, $\theta=0^{\circ}$) and grazing incidence (GI, $\theta=60^{\circ}$); in both cases the magnetic field was collinear with the incident x-rays. Background spectra in the energy range of interest were acquired for each NaCl/metal-substrate sample prior to the deposition of RE atoms, and were subtracted from the RE/NaCl/metal-substrate spectra to eliminate any contribution from the substrate. The magnetization curves were acquired by recording the maximum of the XMCD intensity as a function of the external magnetic field $B$. The photon flux was measured with a photodiode placed after the last optical element of the beamline.

\subsection{NaCl film thickness determination for x-ray measurements}

\begin{figure}[H]
\begin{center}
\includegraphics[scale=0.9]{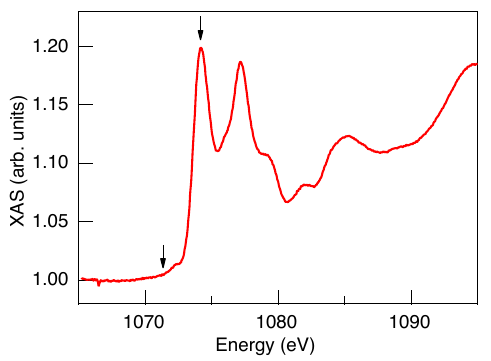}
\caption{Normalized XAS spectrum acquired on 9\,ML of NaCl on Cu(111) at the Na K-edge (room temperature, $B = 0.1$~T). Arrows indicate the energy positions used to determine the Na K-edge height.
\label{nacl}
}
\end{center}
\end{figure}

We calibrated the NaCl coverage combining RT-STM measurements and XAS measurements at the Na K-edge for NaCl on Cu(111). These STM measurements were carried out at the X-Treme end-station. We prepared a sample displaying 25\,\% of its surface covered by NaCl islands. The NaCl growth starts with an initial double layer with smaller regions of additional layers on top of it~\cite{rep04b}, a behavior confirmed in our STM measurements. These STM measurements were used to calibrate the jump at the Na K-edge in terms of absolute NaCl thickness. For thicker NaCl films, we then determined the number of NaCl layers from the intensity of the Na K-edge. Figure~\ref{nacl} shows a typical XAS spectrum acquired on a NaCl/Cu(111) film. This film  has a nominal NaCl coverage of 9\,ML. For NaCl grown on Ag(111), the different absorption and penetration depth of the x-rays in Ag \textit{vs.} Cu has been taken into account.  

\section{D\lowercase{y} atoms: Sample morphology and structural stability}

\begin{figure*}[t]
\includegraphics[scale=1]{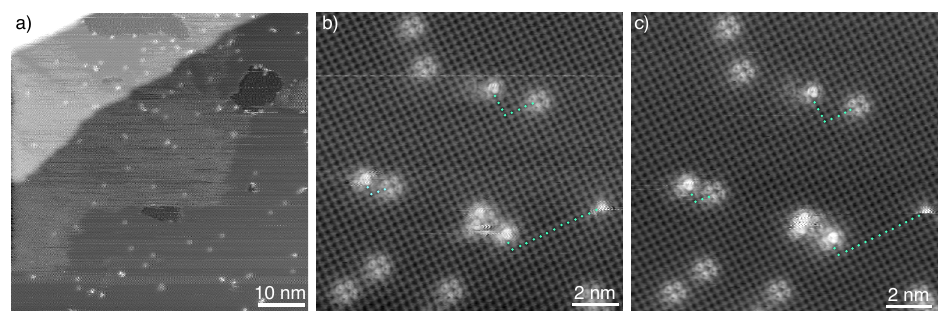}
\caption{Substitutional Dy atoms on NaCl/Ag(100) ($T_{\rm dep} = 300$\,K, $\theta_{\rm {Dy}}=0.4$\,\%\,ML). a) Overview STM image showing individual atoms at the surface. b,c) Consecutive images acquired on the same region demonstrating the absence of atom diffusion during the 10 minutes time-lapse between the two images. ($V_{\rm t}=+200$\,mV, $I_{\rm t}=80$\,pA, $T_{\rm STM} = 300$\,K).
\label{stmrt}
}
\end{figure*}

\subsection{Na-substitutional Dy atoms}
\label{Dysubstit}
The observation of isolated Dy atoms in the STM images upon RT deposition, although measured at LT, is a strong indication of their structural stability, \textit{i.e.} absence of diffusion, also at RT. If the substitutional Dy atoms were to diffuse at RT, one would expect the formation of clusters, which we do not observe. In principle, cluster formation can be inhibited by a strong repulsion between the atoms. To do so at RT, however, a repulsive barrier of much more than 1\,eV would be needed. For example, for Dy adatoms on graphene, a charge transfer of almost 1\,e$^-$, corresponding to a barrier of the order of 0.5\,eV, generates a repulsion inhibiting dimer formation below 100\,K ~\cite{piv18}. For a similar charge transfer for substitutional Dy atoms on NaCl, this barrier would readily be overcome at RT and therefore clusters were formed if the atoms diffused. 

In order to assess diffusion at RT directly, we have carried out STM measurements at RT. We have imaged the sample for several days after deposition of Dy at RT, and with the sample kept at RT all the time. The large scale STM image in Fig.~\ref{stmrt}(a) shows a sample morphology that is qualitatively equivalent to the one observed in LT measurements, \textit{i.e.}, there are small protrusions corresponding to the individual substitutional Dy atoms. This demonstrates that the substitutional atoms do not form islands or clusters.
Nevertheless, they could hop from time to time to the neighboring site. In order to detect whether such motion exists, therefore, we have acquired zoomed-in images displaying the atomic NaCl lattice. The tunneling conditions used for these measurements pull the Dy atoms  out of their subtitutional sites, leaving vacancies that are visible as groups of four brighter Cl atoms, see Fig.~\ref{stmrt}(b). Some substitutional atoms are still present, and appear as brighter features (although slightly perturbed by the tip-sample interaction). The observation, in consecutive images, of unchanged distances (in atomic sites, green dots) between Dy atoms and vacancies or between two Dy atoms demonstrates that the Dy atoms do not diffuse, at least during the timelapse of these measurements (10 minutes between the two images). 
This finding is in line with results reported for Br$^-$ ions replacing Cl$^-$ ions in the surface layer of NaCl, displaying remarkable stability at RT~\cite{kaw14}. The authors have shown that the Br$^-$  ions can be manipulated by using specific conditions in AFM measurements, but that for standard imaging conditions the Br$^-$ ions do not diffuse on the timescale of the experiments.

Finally, energy considerations from DFT also support the structural stability of the substitutional atoms. We have estimated the energy difference between the substitutional RE configuration and a transient configuration during the hopping process, consisting of the Na-vacancy left behind on the substrate and the RE adatom. In the 6\,ML NaCl slab, where the top-Cl is the lowest-energy transient site, the energy differences with respect to the Na-substitutional configuration are 3.56 and 3.92\,eV for Gd and Eu, respectively. For the 2\,ML/Ag(100) NaCl film, the lowest-energy transition state is the bridge site. Here the energy differences are 4.53 and 4.67\,eV for Gd and Eu, respectively. These energies can't be overcome at RT and confirm the stability of the substitutional atoms against diffusion processes (see Sec.~\ref{DFT} for details on the DFT calculations).

\subsection{Dy adatoms}

We have investigated the thermal stability of the Dy adatoms by Dy deposition at different temperatures. Figure~\ref{adat_temp} shows STM images acquired at $T_{\rm STM} = 6.0$\,K for Dy deposition at a) $T_{\rm dep} = 10$\,K,  b) $T_{\rm dep} = 33$\,K, and c) $T_{\rm dep} = 55$\,K on a sample exposing 2\,ML and 3\,ML NaCl regions. In Fig.\ref{adat_temp}(a) there is coexistence of small and larger protrusions, with the small ones being most abundant on the 2\,ML part while the larger ones are the majority species on the 3\,ML part. The small objects, with apparent heights between 30 and 100\,pm, are identified as substitutional Dy atoms ({LT-substitutional), however, they are different from the ones obtained upon RT deposition. In fact, their appearance is less homogeneous, as shown in Fig.~\ref{substLT} and discussed in Sec.~\ref{LTsubtit}. The larger features in Fig.\ref{adat_temp}(a), with an apparent height of the order of 330\,pm,  are Dy adatoms. 
For deposition at 33\,K, shown in Fig.\ref{adat_temp}(b), Dy adatoms are still present on the 3\,ML NaCl regions, while on the 2\,ML region only LT-substitutional species are observed. For deposition at 55\,K, in Fig.\ref{adat_temp}(c), no adatoms are observed on any of the two thicknesses. From the observed trend, adatoms could be present at this temperature on thicker NaCl films.  

%

\begin{figure*}[htbp]
\includegraphics[scale=1]{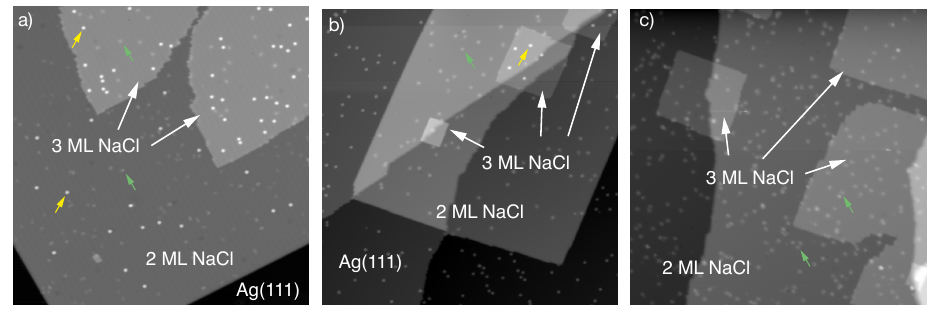}
\caption{Temperature and NaCl thickness dependent adsorption sites  on NaCl/Ag(111): a) $T_{\rm dep} = 10$\,K, $\theta_{\rm {Dy}}=0.2$\,\%\,ML;  b) $T_{\rm dep} = 33$\,K, $\theta_{\rm {Dy}}=0.2$\,\%\,ML  c) $T_{\rm dep} = 55$\,K, $\theta_{\rm {Dy}}=0.3$\,\%\,ML. Yellow arrows indicate examples of adatoms, green arrows examples of LT-substitutional atoms. For all images, 135\,nm $\times$ 135 nm,   $V_{\rm t} = +300$\,mV,  $I_{\rm t} = 50$\,pA, $T_{\rm STM }= 6.0$\,K. 
\label{adat_temp}
}
\end{figure*}

Figure~\ref{stm_adat}(a) shows Dy adatoms on 3\,ML NaCl/Ag(100), obtained upon low-temperature ($T_{\rm dep}= 10$\,K) deposition.  The uniform adatom apparent height of $(335 \pm 25)$\,pm indicates a unique  adsorption site. 
To determine it, we have used the indirect technique already applied to other systems for which it is not possible to obtain atomic resolution on the support without displacing the adatoms with the STM tip. It consists in overlaying a grid or an actual atomic resolution image onto the adatom image; care must be taken to ensure that both images are free from creep or drift artifacts~\cite{fer17, fer17thesis}. The Dy adatoms are found on bridge sites, as shown in Fig.~\ref{stm_adat}(b). This result also holds for 2 and 4 ML thick NaCl films. 
 This observation, however, is fully compatible with the existence of top-Cl adatoms on thicker NaCl films, since the energetically favorable adsorption site can depend on the film thickness. For instance, on a similar system, MgO/Ag(100), the proportion of top-O and bridge adatoms depends on the MgO thickness, as reported not only for Dy but also for other rare earths~\cite{don16, don21, fer17thesis}. From our results, we can deduce that in the NaCl case the minimum film thickness to obtain top-Cl Dy adatoms is larger than 4\,ML. STM measurements on such NaCl films or even thicker, as the ones used in the XAS-XMCD experiments, are not possible using tunneling conditions that do not displace or remove the adatoms because of the NaCl insulating nature. 

An additional difference between the sample preparation for XMCD and STM measurements that might explain the absence of top-Cl adatoms in our STM data is the Dy deposition temperature ($T_{\rm dep}= 4$\,K for XMCD, $T_{\rm dep}= 10$\,K for STM). This parameter can also play a role in the population of the adsorption sites, as shown for Ho on MgO~\cite{fer17thesis}.

\begin{figure}[htbp]
\includegraphics[scale=1]{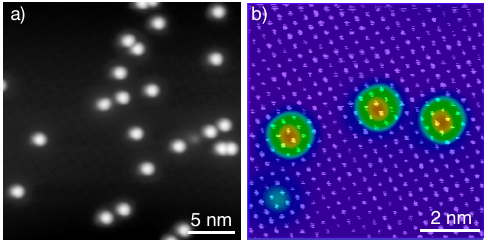}
\caption{Dy adatoms on 3\,ML NaCl/Ag(100) ($T_{\rm dep} = 10$\,K, $\theta_{\rm Dy}= 0.5$\,\%\,ML). a) Overview STM image ($V_{\rm t}=+300$\,mV, $I_{\rm t}=30$\,pA, $T_{\rm STM}= 6.0$\,K). b) STM image showing three adatoms, with superimposed atomic resolution image showing the Cl atoms as protrusions ($V_{\rm t}=+12$\,mV, $I_{\rm t}=100$\,pA) revealing the bridge adsorption site. The very few dim features visible in both images are LT-substitutional Dy atoms.
\label{stm_adat}
}
\end{figure}

\section{LT-substitutional D\lowercase{y}  atoms }
\label{LTsubtit}

 The substitutional atoms obtained upon LT deposition shown in Fig.~\ref{substLT}(a) have a broad distribution of apparent heights between 30\, and 100\,pm, and irregular shapes (asymmetric lateral extension), while the ones obtained upon RT deposition have a very homogeneous appearance, as visible in Fig.~\ref{substLT}(b) as well as in Fig.~1 of the main text. It is possible that LT-substitutional Dy atoms form by knock-on exchange, where part of the adsorption energy is used to overcome the activation barrier for substitution~\cite{kel96, rae96}. The fact that their appearance is different from the RT-subsitutional Dy atoms might be ascribed to out-of-equilibrium geometry, local NaCl distortion, and the presence of replaced atoms in their proximity. 

 X-ray absorption spectroscopy characterization has been carried out on a sample with nominal NaCl thickness of 1.2\,ML with Dy deposition at $T_{\rm dep}= 4$\,K. Since the NaCl growth starts with a double-layer thickness~\cite{rep04b}, roughly 60\% of the surface is covered with NaCl, and the remaining is bare Ag(111). Dy adatoms on Ag(111) have $4f^{10}$ electronic configuration~\cite{sin17}; their XAS spectra have been acquired on a clean Ag(111) surface and subtracted, according to the bare surface fraction, from the 1.2\,ML XAS to obtain the spectra of the LT-substitutional atoms. Figure~\ref{substLT}(c) shows XAS and XMCD spectra, obtained with this procedure, at NI and GI, together with the spectra already shown in Fig.~1 of the main text for RT-substitutional atoms for a direct comparison. From these spectra we deduce that LT-substitutional have a $4f^9$ electronic configuration, small anisotropy energy and a close-to in-plane easy magnetization axis, contrary to the RT-substitutional atoms that are strongly out-of-plane. The weak anisotropy is also reflected by the smaller XLD signal shown in Fig.~\ref{substLT}(d).

 The magnetization curves for the LT-substitutional Dy atoms are acquired at the energy corresponding to the main XMCD peak of the $4f^9$ configuration. The NI and GI $M(B)$-curves, shown in Fig.~\ref{substLT}(e), are both hysteretic, with a narrow opening for $|B| < 1$\,T. This means that LT-substitutional Dy atoms possess long spin relaxation time. However, since the structure of these species is not uniform, we haven't attempted DFT calculations to find the charge distribution around the atoms to determine the CF for the multiplet simulations. 

\begin{figure*}[htb]
\includegraphics[scale=1]{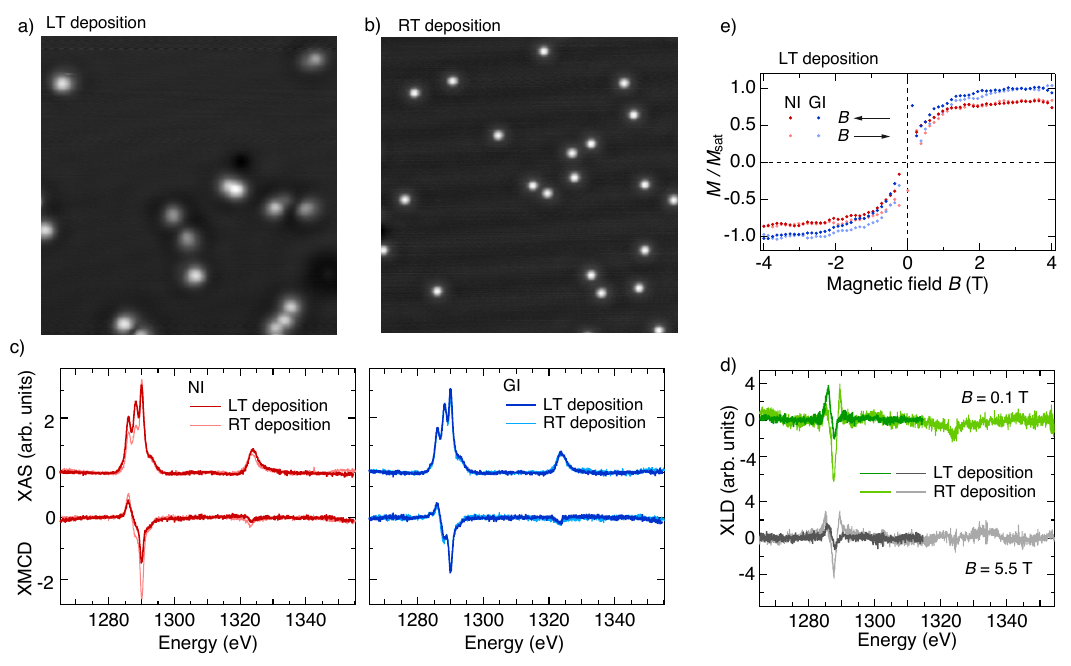}
\caption{Comparison between LT- and RT-substitutional Dy. a) LT-substitutional Dy atoms on 2\,ML NaCl/Ag(111) ($T_{\rm dep}= 10$\,K, $\theta_{\rm {Dy}}=0.2$\,\%\,ML); b) substitutional Dy atoms on 2\,ML NaCl/Ag(111) ($T_{\rm dep}= 300$\,K, $\theta_{\rm {Dy}}=0.5$\,\%\,ML); for both images: 22\,nm $\times$ 22\,nm, $V_{\rm t}=+100$\,mV, $I_{\rm t}=60$\,pA, $T_{\rm STM}= 6.0$\,K, and same grey scale. Comparison of  x-ray absorption results for LT-substitutional Dy atoms on 2\,ML NaCl/Ag(111) ($T_{\rm dep}= 4$\,K, $\theta_{\rm {Dy}}=2.4$\,\%\,ML) and the RT-substitutional ($T_{\rm dep}= 300$\,K, $\theta_{\rm Dy}=1.5$\,\%\,ML) Dy atoms already shown in Fig.\,1 of the main text. c,d) XAS, XMCD  spectra at NI and GI ($B=5.5$\,T), and XLD spectra for LT-substitutional Dy atoms, RT-substitional for comparison; e) magnetization curves of LT-substitutional Dy atoms acquired by recording the XMCD signal at 1289.3\,eV  at NI and GI ($|\dot B | = 2$\,T\,/min, $\phi =1.2\cdot10^{-2}$\,ph\,nm$^{-2}$\,s$^{-1}$). All x-ray absorption data are acquired at $T= 2.5$\,K. 
\label{substLT}
}
\end{figure*}

\section{Absence of magnetic interactions between D\lowercase{y} atoms}
\label{coupling}

The NaCl(100) surface unit cell area is $\approx 0.16$\,nm$^2$. Therefore, a Dy coverage of 1.9\,\%\,ML, used for the  XAS-XMCD measurements reported in Fig.~2 of the main text, corresponds to an average distance between Dy atoms of about 3\,nm. At these interatomic distances, the dipole-dipole interaction between spins is negligible. For instance, the magnetic field generated by a top-O Dy adatom on MgO(100) has been measured and amounts to 0.3\,mT at a distance of 3\,nm~\cite{sin21}. 
In addition to dipolar interactions, there might be exchange interactions. However, also these are expected to be negligible. Superexchange coupling can be disregarded since it requires  hybridization between the RE and NaCl states. NaCl is an ionic material and the RE-Na and RE-Cl bonds are also essentially of ionic character (there are only tiny localized covalent contributions from the RE(d) orbital, as shown in Fig.\ref{fig:pdos_gd}). This excludes superexchange since the needed hybridization is insufficient in our system. 
In the ultrathin NaCl limit, the Ag substrate could contribute to a RKKY exchange. In the DFT calculations we do not observe a spin-polarization of the interface, neither of the Ag layers, induced by the RE atom. Therefore, also RKKY coupling is expected to be negligible. 

At the used coverage, statistical formation of dimers or very small clusters is unavoidable.  Although no detailed investigation of the coverage dependence has been carried out, the STM results are compatible with the data expected for statistical growth~\cite{bru99, leh10, pac15}. This means that at the coverage used in the XAS-XMCD measurements, 1.9\,\%\,ML, approximately 90\,\% of the objects are monomers, the remaining 10\,\% being mainly dimers. In addition, the argument used for the dipolar coupling between adatoms holds also for the coupling between adatoms and dimers or very small clusters, since the expected magnetic field generated by these small clusters cannot be much larger than the one generated by the monomers, \textit{i.e.} a few mT, that are negligible with respect to the applied external field. Note also that dominant clustering would lead to a predominance of $4f^{9}$ electronic configuration, which we do not observe.

\section{Additional X-ray data for samples presented in the main text}
\label{additionalxray}

\subsection{XLD spectra }

Figure~\ref{XLD} shows the XLD spectra acquired for the samples of Fig.~1 (Na-substitutional Dy atoms) and of Fig.~2 (Dy adatoms) of the main text. These spectra are simulated by multiplet calculations with the same CF parameters used for the simulations of the respective XAS, XMCD and magnetization curves. The agreement between the experimental and simulated spectra is satisfactory.

\begin{figure}[htb]
\centering
\includegraphics[scale=1]{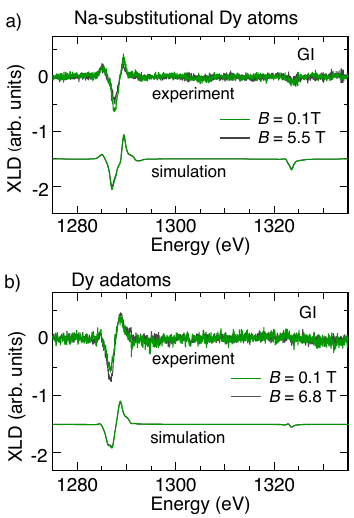}
\caption{XLD spectra acquired for the samples of Fig.\,1 (Na-substitutional Dy atoms) and of Fig.\,2 (Dy adatoms), and simulated spectra.
\label{XLD}
}
\end{figure}

\subsection{Spin relaxation time $T_1$ of Dy adatoms}

In order to characterize the low-field spin dynamics, we acquired the spin lifetime at different fields in NI. In this geometry, more than 95\,\% of the signal comes from top-Cl Dy, as deduced from the angular dependence shown in Fig.~3 of the main text. To perform such measurements, the $B$ field is first set to $6.8$\,T to saturate the sample magnetization and then reduced to the target value, where the difference in absorption at edge (1287.2\,eV) and pre-edge (1280.0\,eV), for a single polarization is acquired as a function of time. Although not being equal to the XMCD signal, this quantity represents the dynamics of the Dy magnetization very well. 

The spin relaxation at $0.3$\,T is shown in Fig.~\ref{tau}(a), together with an exponential fit yielding $T_1 = (550\ \pm\ 100)$\,s. We do not observe significant variations of $T_1$ in the 0.1\,T to 0.5\,T magnetic field range, as shown in Fig.~\ref{tau}(b). This behavior is attributed to scattering with secondary electrons, therefore the measured values are flux limited and represent a lower limit of the intrinsic lifetime.

\begin{figure}[htb]
\centering
\includegraphics[scale=1]{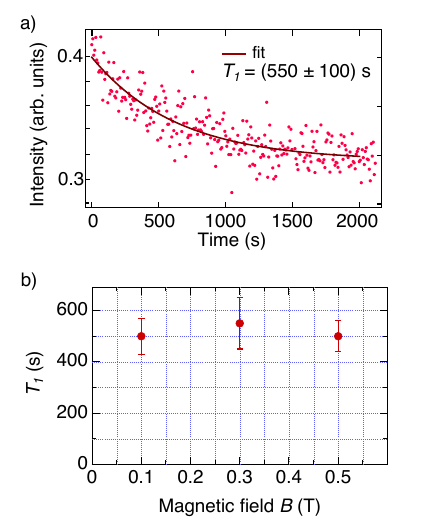}
\caption{$T_1$ measured for the sample of Fig.\,2 of the main text (Dy adatoms). a) Decaying of the difference in absorption between the  M$_5$ edge (1287.2\,eV) and pre-edge (1280.0\,eV) at $B = 0.3$\,T and $T = 2.5$\,K ($\phi = 1 \cdot 10^{-2}$\,ph\,nm$^{-2}$s$^{-1}$\,) and exponential fit yielding $T_1$; b) $T_1$ values vs. $B$ field.
\label{tau}
}
\end{figure}

\section{Density Functional Theory calculations}
\label{DFT}

\subsection{Methodology}

The spin-polarized DFT calculations are carried out in the projector augmented wave approach with plane wave basis sets, as implemented in VASP~\cite{bib:kresse93,bib:kresse99}, using van der Waals (vdW) functionals to address the weak interactions between the RE atom and the NaCl film. The functionals used in this work are based on the DFT-D3 method of Grimme with Becke-Jonson damping 
[D3(BJ)]~\cite{bib:d3bj} and optPBE by Klime{\v{s}} {\it et al.} \cite{bib:optpbe}.

We consider two types of substrates. The first one is a slab of 6 monolayers (ML) of NaCl(100) with the experimental lattice constant 4.085\,{\AA}.  The second type is a heterostructure consisting of a thin film of NaCl(100) on 3\,ML Ag(100), with the NaCl strained to a lattice constant 3.988\,{\AA} to match the Ag lattice. We study heterostructures of  $n=2$ and $3$\,ML films of NaCl(100) on 3\,ML Ag(100). The heterostructure model is based on previous work~\cite{bib:pivetta05},  where the NaCl and Ag lattices are rotated by $45^\circ$, minimizing the mismatch between lattice constants. 
In principle, this can result in different registries between the lattices, where the NaCl atoms fall at hollow or top Ag(100) positions. To minimize interaction between adjacent RE adsorbates, we use supercells of $c(2\times2)$ periodicity. Unless otherwise stated, in this work we focus on the results for the hollow registry and the D3(BJ) functionals.

The convergence thresholds for the total energy and the forces on the atoms are $10^{-7}$\,eV and 0.02\,eV{\AA}$^{-1}$, respectively. Partial state-occupancies in the self-consistency cycles are allowed with the Bl\"ochl triangular method \cite{bib:blochl94} and a 0.1\,eV smearing width for the Fermi level ($E_F$). The plane waves are constructed with a $9\times9\times1$ sampling of the
first Brillouin zone \cite{bib:monk76} and a 280\,eV cut-off.

Na($2p^6 3s^1$), Cl($2s^2p^5$), Gd($5p^66s^25d^1$) and Eu($5p^66s^25d^0$)
are treated as valence electrons in the calculations ($Z_{\rm vcp}$, valence charge of the pseudopotential).
In these systems the RE atoms feature a half-filled
$4f$ orbital, which is inert, thus behaving as Gd$^{3+}$ and Eu$^{2+}$ species.
This allows us to study the effect of the valency on the adsorption
independently of the multiplet effects.
The inclusion of the $4f$ orbital in the valence band is not expected to 
alter the adsorption geometries and energies significantly.
In other words, we expect other RE adatoms with a given valency to show 
similar binding properties, and therefore we take Gd$^{3+}$ and Eu$^{2+}$ as 
representatives of other trivalent and divalent RE species.
Intra-atomic electronic structure effects, such as spin-orbit coupling (SOC),
multiplet splitting due to crystal field and, $f \to d$ electron promotion are disregarded in the present model.

\subsection{Defect formation at the substrates}

To investigate the possibility of RE adsorption at defects,
we create slabs that account for the so-called  Schottky defects,
consisting of a Na and a Cl vacancy. To calculate the defect formation energies,
we use the total energy $E_b$ of a NaCl bulk unit cell with one formula unit 
to set a common reference for the pristine and defective substrates.
For the NaCl 6\,ML substrate we define the Schottky pair formation energy as
\begin{equation}
E_S^{(6L)} = E(\mathrm{NaCl}^{\star\dagger}) + E_b - E(\mathrm{NaCl})
\label{eq:ES_6l}
\end{equation}
where $E(\mathrm{NaCl})$ is the total energy of each 6\,ML slab and the $^{\star\dagger}$ symbols are used to
indicate a pair defect in the slab, consisting of a Na vacant site (Na-vac$^{\star}$) at the top and
a Cl vacant site (Cl-vac$^{\dagger}$) at the bottom of the slab.
In the case of the NaCl n\,ML, with $n=2,3$, films on Ag(100), the Schottky pair formation energy is defined as
\begin{equation}
E_{S,\mathrm{Ag}}^{(nL)} = E(\mathrm{NaCl}^\star\mathrm{/Ag}) + E(\mathrm{NaCl}^\dagger\mathrm{/Ag}) + 
E_b - 2 E(\mathrm{NaCl/Ag})
\label{eq:ES_ag2l}
\end{equation}
where the $^\star$ and $\dagger$ symbols also indicate the presence of Na-vac and Cl-vac sites at
the surface. 

\begin{table*}[btp]
\caption{Formation energies of Schottky pairs in eV for the NaCl films. For the supported NaCl layers, the  values are for NaCl in hollow (top) registry with respect to Ag(100). As reference, the last two columns show the total energy differences between the corresponding top and hollow registries of the pristine films.}
\renewcommand{\arraystretch}{1.8}
\begin{ruledtabular}
 \begin{tabular}{c  c  c  c  c  c}
   vdW functional  &  $E_S^{(6L)}$ &  $E_{S,\mathrm{Ag}}^{(3L)}$  & $E_{S,\mathrm{Ag}}^{(2L)}$ &  $\Delta E^{(2L)}$  & $\Delta E^{(3L)}$\\
\hline
D3(BJ) & 3.525 & 3.267 (3.298) & 2.087 (2.255) &  0.062 &  0.080 \\
optPBE & 3.281 & 3.170 (3.174) & 2.108 (2.189) & -0.043 & -0.042 \\
 \end{tabular}
\end{ruledtabular}
\label{tab:ES}
\end{table*}

The obtained vacancy pair formation energies are shown in Tab.~\ref{tab:ES}. The $E_{S,\mathrm{Ag}}^{(2L)}$ energies are close to the Schottky pair formation energy reported for bulk NaCl in the literature, 2.12\,eV \cite{bib:dreyfus62},  while defect formation is significantly less favored on the NaCl 6\,ML slab and on the 3\,ML NaCl/Ag(100) heterostructure, with differences of $\simeq 1$\,eV with respect to $E_{S,\mathrm{Ag}}^{(2L)}$. Finally, we also find that defect formation is slightly more favorable in the hollow registry than in the top one for 2\,ML, and that this energy difference decreases when the third ML is added. This shows that the defect creation and adsorption of species on the thin NaCl films is weakly sensitive to the structural details of the NaCl/Ag interface. Therefore, we can expect a similar behavior for REs adsorbed on NaCl(100)/Ag(111) and NaCl(100)/Cu(111) films.

\subsection{RE adsorption: structure and stability}
\label{sec:dftenerg}
\begin{figure*}[htb]
\centerline{\includegraphics[width=0.7\textwidth]{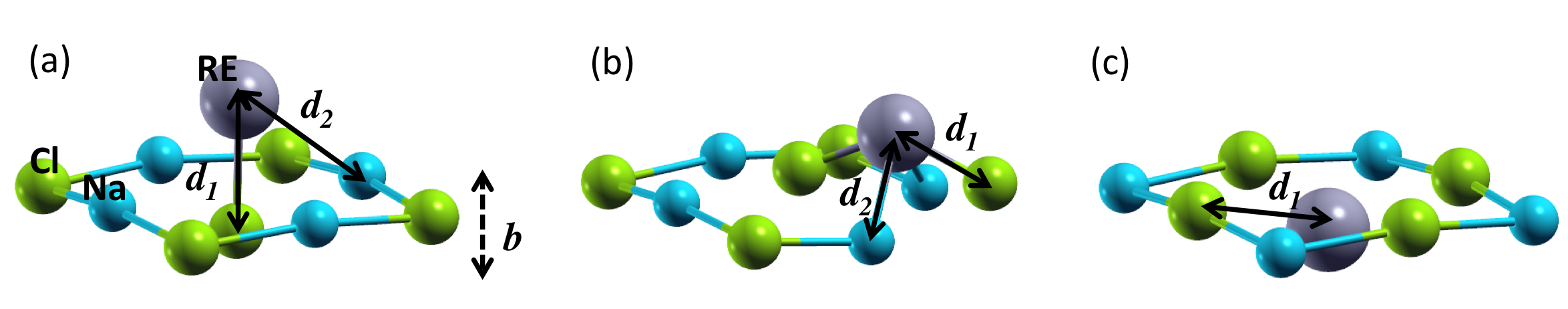}}
\caption{Sketch of the atoms and bonds of RE adsorption with the definitions of the structural quantities shown in Tab.~\ref{tab:DeltaE_geom_6l}, \ref{tab:DeltaE_geom_ag3l}, and \ref{tab:DeltaE_geom_ag2l}. The bonds are artificially distorted and bucklings exaggerated to guide the eye. Color code: Cl in green, Na in blue and RE in gray. Definition of the distances  $d_1 = d (\mathrm{RE-Cl}) $, $d_2 = d (\mathrm{RE-Na}) $ and $b$ (buckling of the topmost NaCl plane) for a RE atom (a) physisorbed at a top-Cl site, (b) at the bridge site of the NaCl lattice, and
(c) substituting a surface Na atom. We note that in the top-Cl and top-Na cases buckling is typically weak, whereas in the bridge case it is stronger, due to the RE-Cl bonding.  In the substitutional case (c) the $b$ value includes the RE atom height and  we use $d_0$ for the RE atom with respect to the average height of the NaCl plane, with $d_0<0$ to indicate that the RE atom lies below such plane.
}
\label{fig:relax_detail}
\end{figure*}

\begin{table}[b]
\caption{
Gd and Eu adsorption energies $\Delta E_{ads}^{(6L)}$ in eV at
the top-Na, bridge and Na substitutional sites on NaCl 6\,ML,
given with respect to the physisorption energy at the top-Cl as reference configuration,
for both vdW functionals.
For the substitutional case, the value $\Delta \tilde E_{ads\star}^{(6L)}$ contains the
energy needed to create a Schottky defect, which are taken from Tab.~\ref{tab:ES}.
The distances in the relaxed structures are in {\AA} (see Fig.~\ref{fig:relax_detail} for definitions).
As reference value, we note that the buckling at clean NaCl(100) is in the 0.09-0.11\,{\AA} range
for the studied functionals, with protruding Cl atoms.
In the bridge site, the RE atom pulls the neighboring Cl atoms outward.
The energetically preferred site is marked in grey.
}
\renewcommand{\arraystretch}{1.7}
\begin{ruledtabular}
 \begin{tabular}{c c     c  c | c c}
         &          & \multicolumn{2}{c|}{Gd}      & \multicolumn{2}{c}{Eu} \\
 site    &          &  D3(BJ)  &   optPBE          & D3(BJ)  &   optPBE     \\
\hline
\colorbox{lightgray}{{ top-Cl}}   & \colorbox{lightgray}{ ${ \Delta E^{(6L)}_{ads} }$}   &   \colorbox{lightgray}{{ 0}}      &  \colorbox{lightgray}{{ 0}}   &   \colorbox{lightgray}{{ 0}}   &   \colorbox{lightgray}{{ 0}}    \\
         &  $d_1$   &  2.59    &  2.63    &   2.76   &  2.82          \\
         &  $d_2$   &  3.87    &  3.91    &   4.11   &  4.16          \\
         &  $b$     & -0.07    & -0.04    &   0.15   &  0.16          \\
\hline
top-Na   &  $\Delta E^{(6L)}_{ads}$   &   0.370  &  0.455  &  0.095   &  0.269    \\
         &  $d_1$   &  4.17   &   4.50    &   4.52   &  4.84           \\
         &  $d_2$   &  3.43   &   6.67    &   3.84   &  4.09           \\
         &  $b$     & -0.35   &  -0.17    &  -0.30   & -0.15           \\
\hline
bridge   &  $\Delta E^{(6L)}_{ads}$  &   0.148  &  0.198  &  0.137   &   0.178   \\
         &  $d_1$   &  2.99   &   3.15    &   3.00   &  3.16           \\
         &  $d_2$   &  3.40   &   3.50    &   3.47   &  3.61           \\
         &  $b$     &  0.28   &   0.24    &   0.32   &  0.34           \\
\hline
subst.   &  $\Delta E_{ads\star}^{(6L)}$  &  -0.036  &  0.176   & -0.395  & -0.213    \\
         &  $\Delta \tilde E_{ads\star}^{(6L)}$           &  3.489   &  3.457   & 3.130  & 3.068    \\
         &  $d_0$   &  0.32   &   0.44    &   0.14   &  0.20           \\
         &  $d_1$   &  2.79   &   2.79    &   2.81   &  2.82           \\
         &  $b$     &  1.39   &   1.16    &   0.17   &  0.21           \\
 \end{tabular}
\end{ruledtabular}
\label{tab:DeltaE_geom_6l}
\end{table}

\begin{table}[b]
\caption{Same information as in Table~\ref{tab:DeltaE_geom_6l} for adsorption on the NaCl 3\,ML
film on Ag(100). }
\renewcommand{\arraystretch}{1.7}
\begin{ruledtabular}
 \begin{tabular}{c c     c  c | c c}
         &          & \multicolumn{2}{c|}{Gd}      & \multicolumn{2}{c}{Eu}  \\
 site    &          &  D3(BJ)  &   optPBE          & D3(BJ)  &   optPBE      \\
\hline
top-Cl   &  $\Delta E^{(3L)}_{ads}$   &  0   &  0    &  0     &  0       \\
         &  $d_1$   &  2.53       &  2.57     &  2.38    &  2.73         \\
         &  $d_2$   &  4.06       &  4.18     &  4.22    &  4.34         \\
         &  $b$     &  0.15       &  0.31     &  0.29    &  0.40         \\
\hline
top-Na   &  $\Delta E^{(3L)}_{ads}$   &  0.649   &  0.796  &   0.266     &  0.465  \\
         &  $d_1$   &  4.08       &  4.62       &  4.53    &   5.18      \\
         &  $d_2$   &  3.40       &  3.81       &  3.83    &   4.48      \\
         &  $b$     &  0.48       &  0.20       &  0.34    &   0.17      \\
\hline
\colorbox{lightgray}{{ bridge}}   &  \colorbox{lightgray}{${ \Delta E^{(3L)}_{ads} }$}  & \colorbox{lightgray}{{ -0.301}}   & \colorbox{lightgray}{{ -0.222}}   &  \colorbox{lightgray}{{ -0.178}}    & \colorbox{lightgray}{{ -0.154}}   \\
         &  $d_1$   &  2.42      &   2.60     &  2.84    &   2.78        \\
         &  $d_2$   &  3.35      &   4.23     &  3.50    &   3.59        \\
         &  $b$     &  2.23      &   1.03     &  2.59    &   0.73        \\
\hline
subst.   &  $\Delta E_{ads\star}^{(3L)}$  & -1.304   &  -1.253   & -2.022    &  -1.949  \\
         &  $\Delta \tilde E_{ads\star}^{(3L)}$            &  1.963   &   1.917   & 1.245     &   1.221   \\
         &  $d_0$   & 0.001      &   -0.16    & -0.10    &  -0.09        \\
         &  $d_1$   &  2.75      &    2.73    &  2.83    &   2.84        \\
         &  $b$     &  0.17      &    0.22    &  0.23    &   0.18        \\
\end{tabular}
\end{ruledtabular}
\label{tab:DeltaE_geom_ag3l}
\end{table}

\begin{table}[h]
\caption{Same information as in Tab.~\ref{tab:DeltaE_geom_6l} for adsorption on the NaCl 2\,ML
film on Ag(100). }
\renewcommand{\arraystretch}{1.7}
\begin{ruledtabular}
 \begin{tabular}{c c     c  c | c c}
         &          & \multicolumn{2}{c|}{Gd}      & \multicolumn{2}{c}{Eu} \\
 site    &          &  D3(BJ)  &   optPBE          & D3(BJ)  &   optPBE     \\
\hline
top-Cl   &  $\Delta E^{(2L)}_{ads}$   &   0      &  0     &  0      &  0       \\
         &  $d_1$   &  2.51    &  2.54    &   2.56   &  2.68         \\
         &  $d_2$   &  4.10    &  4.25    &   4.25   &  4.40         \\
         &  $b$     &  0.22    &  0.43    &   0.35   &  0.54         \\
\hline
top-Na   &  $\Delta E^{(2L)}_{ads}$   &   0.701  &  0.919 &  0.559  &  0.582   \\
         &  $d_1$   &  4.00    &  4.13    &   4.05   &  4.17         \\
         &  $d_2$   &  3.44    &  3.53    &   3.47   &  3.58         \\
         &  $b$     &  0.60    &  0.53    &   0.57   &  0.51         \\
\hline
bridge   &  $\Delta E^{(2L)}_{ads}$  &  -0.663  & -0.513 & -0.493  & -0.353   \\
         &  $d_1$   &  2.55    &  2.57    &   2.68   &  2.72         \\
         &  $d_2$   &  3.49    &  3.49    &   3.66   &  3.73         \\
         &  $b$     &  1.37    &  1.15    &   0.98   &  0.99         \\
\hline
\colorbox{lightgray}{{ subst.}}   &  $\Delta E_{ads\star}^{(2L)}$  &  -3.105  & -2.830 & -3.078  & -2.901   \\
         &  \colorbox{lightgray}{${ \Delta \tilde E_{ads\star}^{(2L)} }$}           &  \colorbox{lightgray}{{ -1.018}}  & \colorbox{lightgray}{{ -0.722}} &  \colorbox{lightgray}{{-0.991}}  & \colorbox{lightgray}{{ -0.793}}   \\
         &  $d_0$   & -0.32    & -0.33    &  -0.12   & -0.12         \\
         &  $d_1$   &  2.69    &  2.71    &   2.82   &  2.84         \\
         &  $b$     &  0.55    &  0.54    &   0.28   &  0.29         \\
\end{tabular}
\end{ruledtabular}
\label{tab:DeltaE_geom_ag2l}
\end{table}

\begin{figure*}[htbp]
\includegraphics[width=0.856\textwidth]{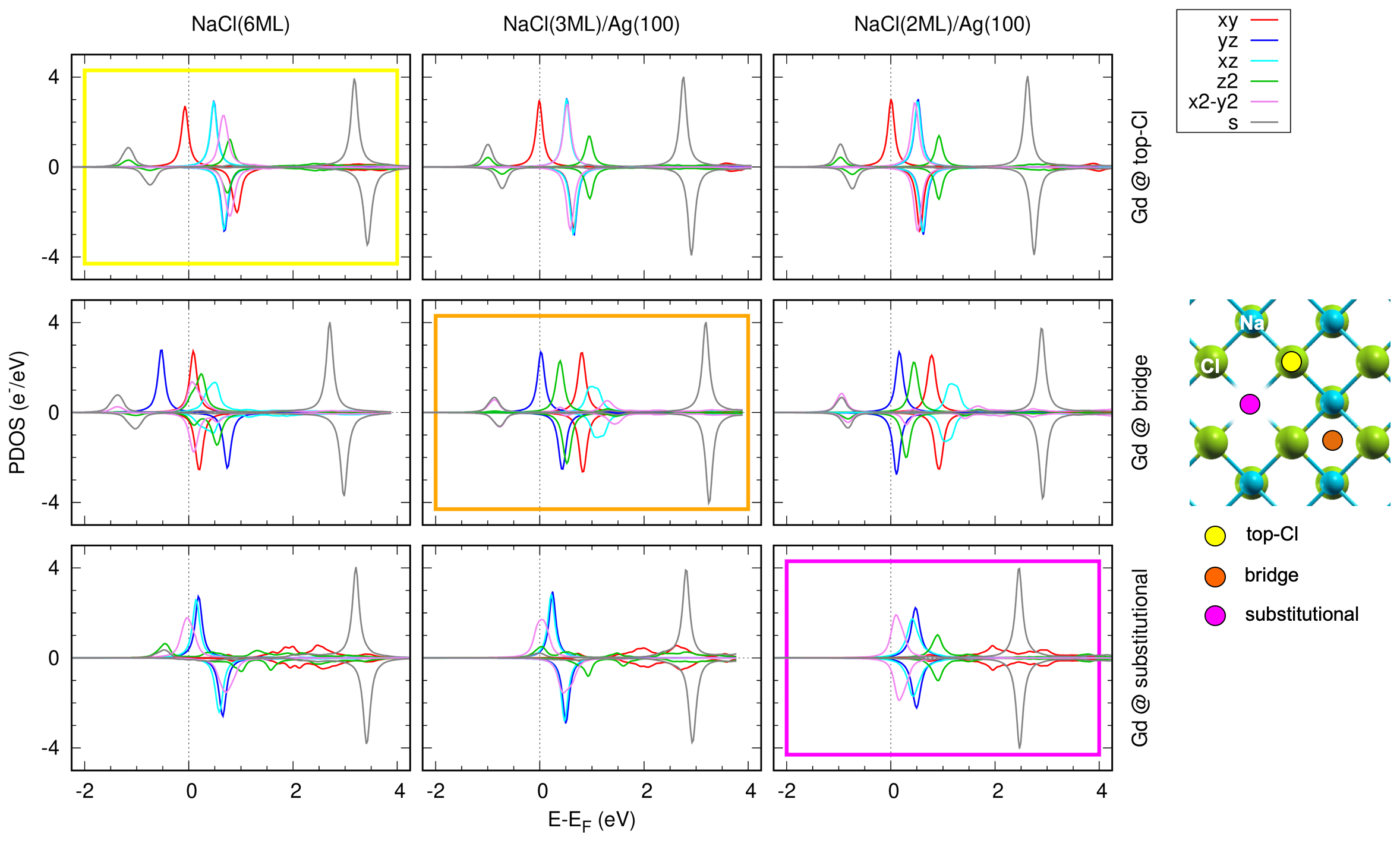}
\caption{Projected densities of states (PDOS) on Gd($s,d$) orbitals, calculated with the D3(BJ) functional for adsorption at different sites (panel rows) of different films (panel columns). The colored rectangles highlight the cases of the energetically preferred  adsorption site for each NaCl film thickness. The sites are depicted in the sketch on the right.
}
\label{fig:pdos_gd}
\end{figure*}

\begin{figure*}[htbp]
\includegraphics[width=0.856\textwidth]{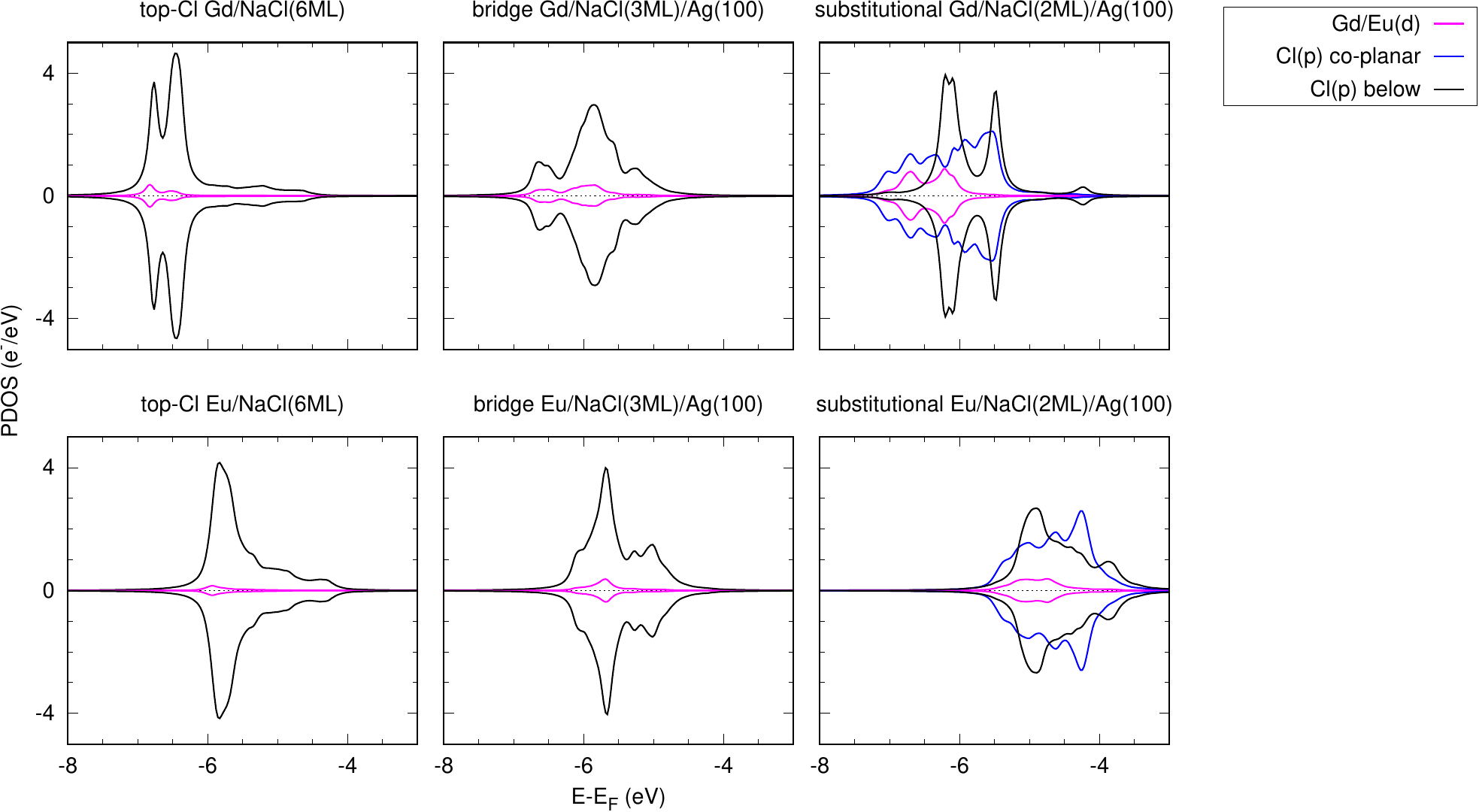}
\caption{Detail of the PDOS on Gd($d$), Eu($d$) and nearest neighboring Cl($p$) orbitals in the preferred adsorption geometries, showing the partially covalent character of the bond between the RE and the lattice.}
\label{fig:pdos_cl}
\end{figure*}


We find local minima in the total energy for Gd and Eu physisorbed at top-Cl and top-Na sites, 
chemisorbed at the bridge sites (hollow between Cl-Cl and Na-Na) and as substitutionals of Na, 
where the RE plays the role of a cation due to the $6s$ electrons 
[see Fig. 4(a,b) of the main paper].
Tab.~\ref{tab:DeltaE_geom_6l}, \ref{tab:DeltaE_geom_ag3l}, and \ref{tab:DeltaE_geom_ag2l}
 show the RE adsorption geometries (see Fig.~\ref{fig:relax_detail} for a description
of the parameters calculated in the structure relaxation), 
and the adsorption energy values at different lattice sites
relative to physisorption at the top-Cl sites, $\Delta E_{ads}$. 
For substitutional RE atoms at Na vacancy sites, 
we first assume that such vacancy exists prior to adsorption of the RE and define
\begin{equation}
\Delta E_{ads\star}^{(6L)} = E(\mathrm{RE/NaCl}^{\star\dagger}) + E_b - E(\mathrm{RE/NaCl})
\label{eq:deltaE_6l}
\end{equation} 
for the NaCl 6\,ML substrate with respect to the top-Cl site energy. 
In the $n$\,ML NaCl/Ag(100) case, with $n=2$ and 3, in order to ensure balance in the atomic species, we define
\begin{multline}
\Delta E_{ads\star}^{(nL)}  = E(\mathrm{RE/NaCl}^\star\mathrm{/Ag}) + E(\mathrm{NaCl}^\dagger\mathrm{/Ag}) +\\ + E_b - E(\mathrm{RE/NaCl/Ag}) - E(\mathrm{NaCl/Ag})
\label{eq:deltaE_ag2l}
\end{multline}

However, the physically meaningful quantity in this case must also include 
the cost of creating the defect in the energy balance, since the STM images
show that there are not point defects on the studied substrates.
In the tables this quantity is defined as 
$\Delta \tilde E_{ads\star}^{(nL)} = \Delta E_{ads\star}^{(nL)} + E_{S,(\mathrm{Ag})}^{(nL)}$.

All species and substrates have preference for a preexisting Na-vac site. 
When the combined process of creating the Na vacant site and replenishing it with the RE is considered, we find 
that the energetically preferred site changes with the film type for both Gd and Eu species:
physisorption at top-Cl for the 6\,ML NaCl slab, 
bridge for 3\,ML NaCl/Ag(100) and substitutional for 2\,ML NaCl/Ag(100).
Nevertheless, it is important to bear in mind that with the present DFT calculations 
we cannot determine whether a divalent or trivalent state will be preferred at
a given adsorption site, since
we cannot compare the total energies of the systems featuring different species and
the intra-atomic $f \to d$ transfer energy contribution is not being considered.
Still, since the Nd, Dy, Ho and Er atoms have similar $f \to d$ promotion
energies of $\simeq 1$\,eV \cite{bib:delin97}, we can argue that we will find them
in similar valency states.

The bridge geometry requires a significant distortion of the NaCl lattice.
In the thin films, the Na-Cl bonds are weakened by the Ag substrate through two effects:
strain on the lattice parameter and modification of the charge density at the NaCl layer in
contact with the metal.
$\Delta E_{ads}$ at bridge is, indeed, reduced by $\simeq 0.3$\,eV when going from 2 to 3\,ML 
for both RE species. At the latter configuration, we also observe that the Gd-Cl bond length is 0.42\,{\AA} 
smaller than the Eu-Cl one, implying a significant role of the $d$-like charge in this particular case.
This is examined in the next subsection.

\subsection{RE adsorption: electronic structure}

\begin{table}[b!]
\caption{Bader charges $q_B$ in units of elementary charge $e$ associated to the RE, Cl and Na atoms 
used to construct the crystal field of $4f^9$ species at the
Na substitutional sites in the 3\,ML NaCl film on Ag(100), calculated with the Gd atom. The $q_B$ values shown here contain the nominal valence charge of the Na and Cl pseudopotentials ($Z_{\rm vcp}=7$), and of the Gd ($Z_{\rm vcp}=9$, $5p^66s^25d^1$).
Distance to the RE $l$ and coordinates $xyz$ with respect to the RE position are given in {\AA}.
}
\renewcommand{\arraystretch}{1.13}
\begin{ruledtabular}
 \begin{tabular}{c  c  c  c  c  c  }
     &  $q_B$ & $l$ &  $x$  & $y$ &  $z$ \\
\hline
Gd &  7.316 & 0.0 &  0.0 &   0.0 &   0.0 \\     

Cl &  7.883 & 2.750 &  1.944 &   1.944 &   0.062 \\     
Cl &  7.880 & 2.750 & -1.944 &   1.944 &   0.062 \\     
Cl &  7.880 & 2.750 &  1.944 &  -1.944 &   0.062 \\     
Cl &  7.887 & 2.750 & -1.944 &  -1.944 &   0.062 \\     

Cl & 7.790  & 2.636 &  0.0   &   0.0   &   -2.636 \\   

Na & 6.130 & 4.089 &  4.085 &  0.0   & -0.110 \\             
Na & 6.130 & 4.089 &  0.0   &  4.085 & -0.110 \\            
Na & 6.130 & 4.089 & -4.085 &  0.0   & -0.110 \\            
Na & 6.130 & 4.089 &  0.0   & -4.085 & -0.110 \\            

Na & 6.128 & 4.228 &  2.105 &   2.105 &   -3.003 \\   
Na & 6.128 & 4.228 & -2.105 &   2.105 &   -3.003 \\   
Na & 6.128 & 4.228 &  2.105 &  -2.105 &   -3.003 \\   
Na & 6.128 & 4.228 & -2.105 &  -2.105 &   -3.003 \\   


\end{tabular}
\end{ruledtabular}
\label{tab:bader_Gd_Navac_3MLAg}
\end{table}

\begin{table}[h!]
\caption{Same as table~\ref{tab:bader_Gd_Navac_3MLAg} for the $4f^{10}$ species at the top-Cl sites of 6\,ML NaCl, calculated with the Eu adatom ($Z_{\rm vcp}=8$, $5p^66s^25d^0$).}
\renewcommand{\arraystretch}{1.13}
\begin{ruledtabular}
 \begin{tabular}{c  c  c  c  c  c  }
     &  $q_B$ & $l$ &  $x$  & $y$ &  $z$ \\
\hline

	Eu &  7.844 & 0.0 &  0.0 &   0.0 &   0.0 \\     
   Cl  &  7.844 &  2.759 &  0.0 &    0.0 &   -2.759 \\   


   Na & 6.135 & 4.112 &  2.054 &   2.054 &  -2.911 \\  
   Na & 6.135 & 4.112 & -2.054 &   2.054 &  -2.911 \\  
   Na & 6.135 & 4.112 &  2.054 &  -2.054 &  -2.911 \\  
   Na & 6.135 & 4.112 & -2.054 &  -2.054 &  -2.911 \\  
 
   Na & 6.129 & 5.827 &  0.0   &   0.0   &  -5.827  

\end{tabular}
\end{ruledtabular}
\label{tab:bader_Eu_topCl_6ML}
\end{table}

\begin{table}[h!]
\caption{Same as table~\ref{tab:bader_Gd_Navac_3MLAg} for the $4f^{10}$ species at the bridge sites of 6\,ML NaCl, calculated with the Eu adatom ($Z_{\rm vcp}=8$, $5p^66s^25d^0$).}
\renewcommand{\arraystretch}{1.13}
\begin{ruledtabular}
 \begin{tabular}{c  c  c  c  c  c  }
     &  $q_B$ & $l$ &  $x$  & $y$ &  $z$ \\
\hline

   Eu & 7.742 & 0.0 &  0.0 &   0.0 &   0.0 \\    

   Cl & 7.936 & 3.003 &  -1.880 &   0.0 &   -2.342 \\    
   Cl & 7.936 & 3.003 &   1.890 &   0.0 &   -2.342 \\    


   Na & 6.158 & 3.470 &  0.0   &  2.223 & -2.665 \\   
   Na & 6.158 & 3.470 &  0.0   & -2.223 & -2.665 \\   


\end{tabular}
\end{ruledtabular}
\label{tab:bader_Eu_bridge_6ML}
\end{table}

In the trivalent Gd case, the adsorption process is to be understood in terms of the transfer and hybridization of the unpaired $d$-electron with the NaCl lattice.  The projected densities of states (PDOS) on the Gd atomic orbitals are shown in Fig.~\ref{fig:pdos_gd} for the NaCl 6\,ML and NaCl 3 and 2\,ML film on Ag.

The cases where peaks are present at the Fermi level or below correspond to configurations where the unpaired $5d$-electron remains mainly localized at the Gd adsorbate. This electron is transferred to the lattice in the case of Gd at bridge and substitutional sites of the thin 2\,ML film, facilitated by the nearby presence of the Ag surface. For Gd physisorbed at top-Cl sites, the state at the Fermi level has $d_{xy}$ character, whereas in the bridge position the unpaired electron occupies the $d_{yz}$ orbital.  The magnetocrystalline anisotropy behavior of these electronic configurations would consist of a contribution to a perpendicular easy magnetization axis in the top-Cl case and to a transversal axis along the Cl-Cl direction in the bridge site case. The small discrepancy between the $d_{xz}$ and $d_{yz}$ orbitals in the Gd substitutional case can be attributed to very weak distortions that make the structure slightly non-symmetric.

There are additional small PDOS peaks with $d$ character in the 
energy interval $[-7,-4]$\,eV for both Gd and Eu species (shown in Fig.~\ref{fig:pdos_cl}).
This peak is contributed by a weak RE($d$)-Cl($p$) hybridization,
which can be interpreted as the bonding between the RE and the NaCl lattice 
not being of pure ionic character, but partially covalent. 

A Bader analysis of the charge densities in the DFT-optimized configurations is used to construct the point charge models used in the multiplet simulations with the Dy atom. Tables~\ref{tab:bader_Gd_Navac_3MLAg}, \ref{tab:bader_Eu_topCl_6ML}, and \ref{tab:bader_Eu_bridge_6ML} show the coordinates and Bader charges $q_B$ for the three configurations presented in Fig.\,4(c,d,e) of the main text. Such figure shows the charge density difference between the self-consistently converged electron distribution and the superposition of atomic (pseudopotential) charges at the atomic positions, \textit{i.e.}, it represents the charge redistribution upon formation of bonds in the system.


\section{Multiplet calculations}

\subsection{Crystal field}

Multiplet calculations of the XAS, XMCD and XLD spectra, as well as of the magnetization curves, are performed using the multiX software, based on effective point charges to compute the CF~\cite{uld12}. To determine these point charges, we apply the procedure developed for Dy and Ho on BaO thin films~\cite{sor23}, that uses the DFT results for the charge distribution around the RE atom. 

In this procedure, the Cl-RE interaction, leading to high interstitial charges, is described by an effective charge whose value ($q_{\rm eff}$) and distance from the RE atom ($R_{\rm eff}$) are given by $q_{\rm eff}=v_{\rm RE}(\epsilon_{\rm Cl}-\epsilon_{\rm RE})/(N\epsilon_{\rm RE})$ and $R_{\rm eff}=R_{\rm Cl}\epsilon_{\rm Cl}/(\epsilon_{\rm RE}+\epsilon_{\rm Cl})$. $v_{\rm RE}=Z_{\rm RE}- q_B$ is the charge contributing to the bond ($q_B$ is the Bader charge calculated by DFT), $N$ the number of nearest Cl neighbors, $R_{\rm Cl}$ the Cl-RE distance, and $\epsilon_{\rm Cl}= 3.16$, $\epsilon_{\rm RE }= 1.23$ (RE = Dy, Ho) the Pauling electronegativity~\cite{zol84}. The Na-RE interaction is represented by a positive charge placed on the Na ionic radius (1.16\,{\AA}) along the Na-RE direction; the value of this positive charge is used as the only fit parameter, with the constraint that it cannot exceed the value deduced from the Bader charge. 
The procedure has been slightly modified for the case of substitutional Dy atoms. Indeed, in order to respect the latter requirement for the positive charges, we had to increase slightly the $q_{\rm eff}$ charge of the axial Cl ion. This change is justified by looking at the cut in Fig.\,4 of the End Matter, showing that the charge relative to the Cl below the RE atom is larger than the one of the Cl atoms in the plane.

The crystal field parameters (point charge coordinates and values) used to simulate the spectra and the magnetization curves shown in the main text for Dy and in SM Sec.~\ref{secHo} for Ho are reported in Tab.~\ref{tabsubst} for Dy substitutional atom, in Tab.~\ref{tabtop} for top-Cl Dy and Ho adatoms, and in Tab.~\ref{tabbr} for bridge Dy and Ho adatoms.

 \begin{table}[h!]
\caption{Point charges used to generate the CF for substitutional Dy atoms: 5 negative charges (nearest Cl atoms), and 8 positive charges (nearest Na atoms).}
  \label{tabsubst}
\renewcommand{\arraystretch}{1.03}
\begin{ruledtabular}
  \begin{tabular}{ c c c c }
   \multicolumn{1}{c}{x, \AA} & \multicolumn{1}{c}{y, \AA} & \multicolumn{1}{c}{z, \AA} & \multicolumn{1}{c}{q, e$^-$} \\
   \hline
   0.00 & 0.00 & -1.87 & -0.675 \\
   1.40 & 1.40 & 0.00 & -0.54 \\
   1.40 & -1.40 & 0.00 & -0.54 \\
   -1.40 & 1.40 & 0.00 & -0.54 \\
   -1.40 & -1.40 & 0.00 & -0.54 \\
   2.92 & 0.00 & 0.00 & 0.87 \\
   -2.92 & 0.00 & 0.00 & 0.87 \\
   0.00 & 2.92 & 0.00 & 0.87 \\
   0.00 & -2.92 & 0.00 & 0.87 \\
   1.53 & 1.53 & -2.16 & 0.87 \\
   1.53 & -1.53 & -2.16 & 0.87 \\
   -1.53 & 1.53 & -2.16 & 0.87 \\
   -1.53 & -1.53 & -2.16 & 0.87 \\
  \end{tabular}
\end{ruledtabular}
\end{table}
\begin{table}[h!]
\caption{Point charges used to generate the CF for top-Cl Dy and Ho atoms: 1 negative charge (nearest Cl atom), and 5 positive charges (4 nearest Na atoms and next-nearest (axial) Na atom).}
  \label{tabtop}
\renewcommand{\arraystretch}{1.03}
\begin{ruledtabular}
  \begin{tabular}{ c c c c }
   \multicolumn{1}{c}{x, \AA} & \multicolumn{1}{c}{y, \AA} & \multicolumn{1}{c}{z, \AA} & \multicolumn{1}{c}{q, e$^-$} \\
   \hline
   0.00 & 0.00 & -1.99 & -0.244 \\
   2.05 & 2.05 & -2.91 & 0.865 \\
   2.05 & -2.05 & -2.91 & 0.865 \\
   -2.05 & 2.05 & -2.91 & 0.865 \\
   -2.05 & -2.05 & -2.91 & 0.865 \\
   0.00 & 0.00 & -5.72 & 0.865 \\
  \end{tabular}
  \end{ruledtabular}
 \end{table}
\begin{table}[h!]
\caption{Point charges used to generate the CF for bridge Dy and Ho atoms: 2 negative charges (nearest Cl atoms), and 2 positive charges (nearest Na atoms).}
  \label{tabbr}
\renewcommand{\arraystretch}{1.03}
\begin{ruledtabular}
  \begin{tabular}{ c c c c }
   \multicolumn{1}{c}{x, \AA} & \multicolumn{1}{c}{y, \AA} & \multicolumn{1}{c}{z, \AA} & \multicolumn{1}{c}{q, e$^-$} \\
   \hline
   0.957 & 0.957 & -1.686 & -0.202 \\
   -0.957 & -0.957 & -1.686 & -0.202 \\
   1.03 & -1.03 & -1.78 & 0.84 \\
   -1.03 & 1.03 & -1.78 & 0.84 \\
  \end{tabular}
    \end{ruledtabular}
 \end{table}

\begin{figure*}[htbp]
\centering
\includegraphics[width=1\textwidth]{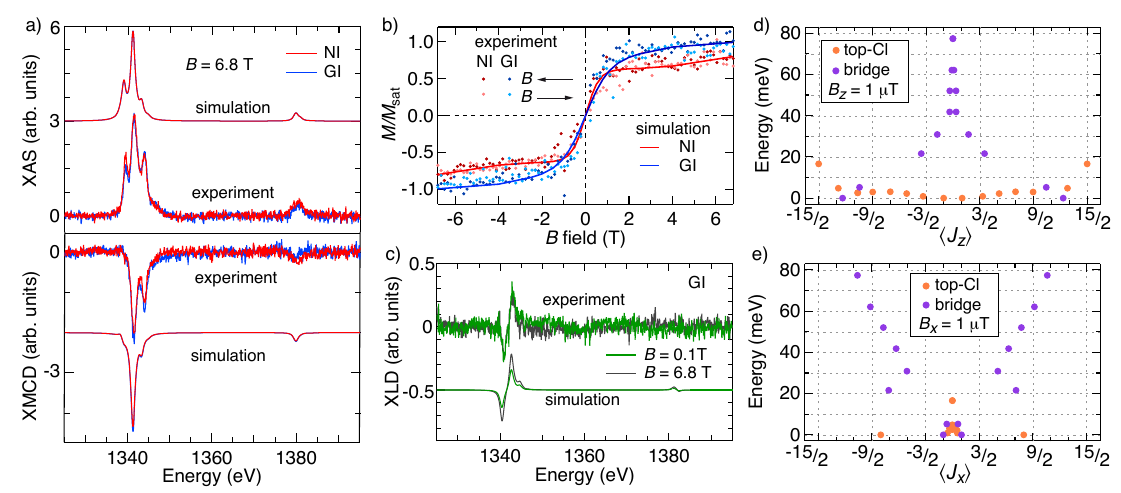}
\caption{Ho  adatoms ($T_{\rm dep}= 4$\,K): $\theta_{\mathrm Ho}=2.1$\,\%\,ML Ho on 8\,ML NaCl/Cu(111). a) XAS and XMCD experimental spectra at the Ho $M_{4,5}$ edges, and simulated ones ($B=6.8$\,T, $T_{\rm XAS}=2.5$\,K). b) NI and GI magnetization curves acquired by recording the XMCD signal at 1342.0\,eV ($|\dot B | = 33.3$\,mT\,s$^{-1}$, $\phi =1.9\cdot10^{-2}$\,ph\,nm$^{-2}$\,s$^{-1}$), and simulated equilibrium curves at $T=2.5$\,K. c) XLD experimental and simulated spectra. d,e) Magnetic level scheme of the ground $J$ multiplet for top-Cl and bridge species. 
\label{Ho}
}
\end{figure*}

\subsection{Simulation of magnetization curves and contribution of $T_z$ to the magnetic anisotropy}

The calculated magnetization curves report the intensity of the XMCD M$_5$ peak of the spectra as a function of $B$ field; this approach replicates the procedure used to measure the experimental curves. For the substitutional  Dy atoms of Fig.\,1 of the main text, the ratio of the saturation magnetization at grazing and normal incidence calculated in this way differs from the experimental one by about 5\%. Although this is a small discrepancy, the amplitude of the simulated curves at 6.8\,T is normalized to the amplitude of the experimental curves to have a better comparison of the field dependence.

According to the sum rules~\cite{tho92, car93}, the XMCD peak, and thus the experimental signal reported in the magnetization curves, is actually proportional to $I(M_5) = (3/2L_z +2S_z + 6T_z)t/(5h)$, where $t$ is the XAS integral and $h=5$ the number of holes in Dy $4f^9$. Thus $I(M_5)$ differs from the actual magnetization. However, in RE and actinides, where the CF is a perturbation with respect to the SOC, the spin density distribution rotates according to the magnetization direction, implying that $\mathbf{L}$, $\mathbf{S}$, and $\mathbf{T}$ are parallel~\cite{ogu04}. Therefore, we expect $I(M_5)$ to represent the field dependence of the magnetization very well, we further expect that the contribution of $T_z$ to $I(M_5)$ is small and that it does produce an artificial magnetic anisotropy.

To verify this statement for our system, we evaluate the contribution of $T_z$ to the magnetic anisotropy. Since multiX does not provide the $T_z$ value, we have used the following approximate procedure to estimate it~\cite{sin17}:
\textit{i}) we have applied the sum rules to the simulated XAS and XMCD spectra to determine $S_{\rm eff}= 2S_z +6T_z$; \textit{ii}) we have determined $T_z$ using this $S_{\rm eff}$ value and the $S_z$ value provided by multiX.

We find that $6T_z= 0.72$ at NI and $6T_z= 0.57$ in GI, in units of $\hbar$. These values correspond to 7.5\% and 10\% of the corresponding $I(M_5)$, meaning that the $T_z$ contribution is basically isotropic. Therefore, the difference in saturation values ($M_{\rm sat}$@GI / $M_{\rm sat}$@NI $ \approx 0.65$) and the low-field susceptibilities
measured at NI and GI are only marginally affected by $T_z$.

\section{H\lowercase{o on} 8 ML N\lowercase{a}C\lowercase{l}}
\label{secHo}

Ho deposition onto NaCl films with nominal thickness of 8\,ML, held at $T=5$\,K, results in an ensemble of Ho adatoms at top-Cl and bridge sites, similar to what we report for Dy in the main text, and to Ho or Dy on various oxides~\cite{don21, bel22, sor23}. The XAS spectra shown in Fig.~\ref{Ho}(a) demonstrate a $4f^{11}$ occupancy~\cite{bal16, sin17, don21}. The peak at 1344.0\,eV reveals a $4f^{10}$ contribution, attributed mainly to contaminated species, since its intensity increases with beam exposure time~\cite{bal16, bal18}. The ensemble displays a canted magnetization, as revealed by the almost identical intensities of the XMCD spectra acquired in NI and GI, and by the saturation values of the magnetization curves shown in Fig.~\ref{Ho}(b). No sign of hysteresis is observed in the magnetization curves, in strong contrast with the behavior observed for top-O Ho on MgO~\cite{don16}. 

To rationalize these results, we use the very same CF applied to Dy, for both top-Cl and bridge Ho adatoms, to perform multiplet calculations. The experimental observations are reproduced in all respects: the XMCD intensity in NI and GI, the shape and saturation value of the magnetization curves in NI and GI, the XLD shape and intensity, together with its weak field dependence. The best agreement is obtained with an ensemble composed of 40\,\% top-Cl and 60\,\% bridge adatoms, slightly different than the proportion used for Dy discussed in the main text. 

The corresponding quantum level schemes are shown in Fig.~\ref{Ho}(d,e). Top-Cl Ho has an easy-plane anisotropy, while bridge Ho has out-of-plane easy axis. The in-plane behavior for top-Cl might appear surprising when compared with the behavior of top-O Ho on MgO~\cite{don16} and on BaO~\cite{sor23}, both showing out-of-plane easy magnetization axis. This difference can be rationalized considering the $4f^{11}$ occupancy on NaCl in contrast with the $4f^{10}$ one on the two oxides. As a consequence, on NaCl the spatial distribution of the total $4f$ charge density resembles the one of Er $4f^{11}$, which is prolate for maximum $J_z$~\cite{rin11}. Indeed, for Er $4f^{11}$ it has been shown that negative charge localized below (and above) the atom promotes a configuration with oblate charge distribution corresponding to a low $J_z$ value as ground state, while lateral charges promote a prolate charge distribution~\cite{rin11, don14, sin16}. Therefore, top-Cl Ho adopts an oblate charge distribution with low $J_z$ as ground state. 

\clearpage

\bibliography{ms_16.bib}